\documentclass[prd,
a4paper,
nofootinbib,
onecolumn,
reprint,
eqsecnum,
]{revtex4}
\usepackage{multirow}
\usepackage{graphicx}
\usepackage{amsmath}
\usepackage{amsfonts}
\usepackage{amssymb}
\usepackage{slashed} 
\usepackage{color}
\usepackage{hyperref}
\usepackage{units}
\usepackage{braket}
\usepackage{siunitx}

\def\Im{\mbox{Im}\,}
\def\Re{\mbox{Re}\,}

\def\eg{{\it e.g.}}
\def\ie{{\it i.e.}}

\definecolor{oucrimsonred}{rgb}{0.6, 0.0, 0.0}
\definecolor{persianblue}{rgb}{0.11, 0.22, 0.73}
\definecolor{forestgreen}{rgb}{0.13,0.35,0.13}
 \hypersetup{colorlinks, citecolor=oucrimsonred, linkcolor=persianblue, urlcolor=oucrimsonred}
\newcommand{\be}{\begin{equation}}
\newcommand{\ee}{\end{equation}}
\newcommand{\bea}{\begin{eqnarray}}
\newcommand{\eea}{\end{eqnarray}}

\newcommand{\into}{\rightarrow}
\begin{document}

\title[]{A Minimal Dark Matter Model for Muon g-2 with Scalar Lepton Partners up to the TeV Scale
}
\date{\today}
\author{Jan Tristram Acu\~{n}a$^{\ast\dag\ddag}$}
\author{Patrick Stengel$^{\ast\dag\ddag}$}
\author{Piero Ullio$^{\ast\dag\ddag}$}
\affiliation{$^{\ast}$Scuola Internazionale Superiore di Studi Avanzati (SISSA), via Bonomea 265, 34136 Trieste, Italy}
\affiliation{$^{\dag}$INFN, Sezione di Trieste, via  Valerio 2, 34127 Trieste, Italy }
\affiliation{$^{\ddag}$Institute for Fundamental Physics of the Universe (IFPU), via Beirut 2,  34151 Trieste, Italy}
\begin{abstract}
\noindent  
The E989 experiment at the Fermi National Laboratory reported a 4.2$\sigma$ discrepancy between the measured magnetic dipole moment of the muon, and its prediction in the Standard Model (SM). In this study, we address the anomaly by considering a minimal and generic extension to the SM which also provides for a dark matter (DM) candidate. The extra states in this framework are: a SM singlet Majorana fermion, referred to as the Bino, playing the role of DM; and muonic scalars, referred to as sleptons. The couplings between the sleptons, SM muons and the Bino can account for the muon $g-2$ anomaly if the scalar muon partners, or smuons, mix chirality. On the other hand, the DM relic density is satisfied primarily through coannihilation effects involving the Bino and the lighter sleptons. The viable parameter space of our model includes regions with relatively light coannihilating particles, similar to what has been found in previous scans of the Minimal Supersymmetric Standard Model (MSSM). Relaxing the assumption of minimal flavor violation typically assumed in the MSSM, we see that scenarios with sizable smuon mixing and large mass splittings between the smuons can satisfy both the muon $g-2$ anomaly and the DM relic density for coannihilating particle masses up to and beyond the TeV scale. When we specify the origin of the left-right smuon mixing to be trilinear couplings between the smuons and the SM Higgs boson, the constraints on these scenarios arising from perturbative unitarity and electroweak vacuum stability confine the coannihilating particle masses to be $\lesssim 1 \,$TeV. We demonstrate that next generation direct detection experiments are only marginally sensitive to the viable parameter space of our model and, thus, a future lepton collider could be the essential probe necessary to distinguish our model from other BSM solutions to the muon $g-2$ anomaly. 
\end{abstract}

\maketitle

\section{Introduction}

The identification of the nature of the dark matter (DM) component of the Universe remains one of most pressing open problems in science today. While cosmological and astrophysical data can provide some insight into the properties of DM, there is no direct evidence that DM corresponds to a new elementary particle (or a new sector of particles). One approach in the last several decades has thus been to look for other hints of new physics which could be related to the DM problem. Besides its numerous successes, the Standard Model (SM) of particle physics is unable to address several questions, including failing to embed candidates for DM and dark energy, and a working mechanism for the generation of the baryon asymmetry in the Universe. While recent searches for the direct production of new particles at the high energy frontier have been unsuccessful, there have been a number of anomalies emerging at the high precision frontier, possibly indirectly pointing to new interaction states. In particular, for the last few years there has been intriguing signs of new physics in several flavor physics anomalies, see, \eg,~\cite{LHCb:2021trn,HFLAV:2019otj}.

One of the longest standing potential anomalies within the SM is the discrepancy between the measured values for the anomalous magnetic moment of the muon and its predicted value. Recently the E989 experiment at the Fermi National Laboratory reported its first results~\cite{Muong-2:2021ojo}, confirming, with higher precision, the picture that had already emerged in 2006 with the final report from the E821 experiment at the Brookhaven National Laboratory \cite{Muong-2:2006rrc}: the measured magnetic anomaly parameter for the muon $a_\mu = (g_\mu -2)/2$ differs from its best up-to-date SM prediction~\cite{Aoyama:2020ynm} at a level which starts to be statistically intriguing, about 4.2$\sigma$ when combining the result from the two experiments~\cite{Muong-2:2021ojo}:
\be
\Delta a_\mu^{exp} = (25.1 \pm 5.9) \times 10^{-10}\,.
\label{eq:deltamuexp}
\ee
While the debate regarding the SM computation of $a_\mu$ and its uncertainty is still ongoing, the discrepancy has attracted significant attention in the last two decades since there are several extensions to the SM in which a sizable contribution to $g_\mu-2$ is predicted (the literature in this respect is vast, see, \eg, the recent reviews and general discussions~\cite{Athron:2021iuf,Jegerlehner:2017gek,Lindner:2016bgg,Jegerlehner:2009ry,Melnikov:2006sr,Stockinger:2006zn}). Matching the anomaly with an extra contribution at 1-loop level is possible in rather generic scenarios; the general requirement is to introduce beyond-the-SM (BSM) states that couple to the muon and/or carry muonic lepton number, flip chirality, and are either electrically charged or participate in mediating another coupling to photons. Most minimal setups, featuring a single new BSM field flowing in the loop diagram, such as a second Higgs doublet~\cite{Broggio:2014mna,Cherchiglia:2016eui}, a leptoquark~\cite{Chakraverty:2001yg}, an axion-like particle~\cite{Marciano:2016yhf}, or a dark photon~\cite{Pospelov:2008zw}/dark Z~\cite{Davoudiasl:2012qa}, have been systematically studied;  in general, they are severely constrained by other observables, see, \eg, the update in~\cite{Athron:2021iuf}, and, most notably from our point of view, they all fail to provide a DM candidate. 

A BSM state can play the role of DM if it fulfills several fairly generic requirements: it is stable or very long-lived, its coupling to photons is very strongly suppressed (and it is color neutral), its self-interactions are not too strong, and it starts driving the gravitational collapse of bound structures at the onset of the matter-dominated epoch (DM must be cold or, at most, warm). A key ingredient is also the identification of a viable production mechanism for this state in the early Universe. Accommodating these features and accounting for the $g_\mu-2$ excess is possible in rather minimal SM extensions, and the goal of this paper is to highlight features of one of these most minimal frameworks. With respect to other cases studied so far, the scheme considered here is interesting from two perspectives. This scenario is at the same time minimal from the point of view of having minimal BSM particle content, as well as being the minimal working recipe within well motivated, more extended frameworks for BSM physics such as generic supersymmetric SM extensions, and the Minimal Supersymmetric SM (MSSM) in particular. 

We consider a setup with at least two BSM fields relevant for $g_\mu-2$ and the DM relic density; we assume that both are involved in the 1-loop diagram providing for the extra contribution to $a_\mu$ and that one of them is neutral and accounts for DM. There are a few different possible choices (see, \eg, the discussion in~\cite{Calibbi:2018rzv}) depending on which of the two is a fermion and which is a boson, which carries muonic lepton number, and how the muon chirality flip proceeds (excluding the possibility that it comes only from a mass insertion on the external legs of the relevant 1-loop diagram). We will focus on the case in which the neutral particle has zero muonic lepton number and is spin 1/2, a Majorana state coupled to both the left-handed muon and the right-handed muon via a charged scalar lepton partner, which carries muon lepton number and mixes chirality. With this particular and peculiar choice, while correctly assigning the $SU(2)_L \times U(1)_Y$ quantum numbers, we are selecting a small subset of the particle content of the MSSM, one of the frameworks in which BSM contributions to $g_\mu-2$ have been first and most extensively studied, with some of the earliest references including, \eg,~\cite{Grifols:1982vx,Ellis:1982by,Barbieri:1982aj,Kosower:1983yw}. In the MSSM jargon, which we will adopt in the rest of the paper, we are considering a scenario with a pure Bino DM candidate, and muon sleptons the only other light (or relevant) supersymmetric partners, hence assuming, \eg, that all other neutralinos and the charginos are very heavy and decoupled. 

Nonetheless, the model we consider is not simply zooming in on a particular case generically included within a MSSM parameter scan: the only contribution to $g_\mu-2$ included here is most often very subdominant. In most realizations of the MSSM, the mixing of right-handed and left-handed sleptons is assumed to be negligible, with the exception of models introduced in~\cite{Fukushima:2014yia} and follow up papers~\cite{Kelso:2014qja,Kumar:2016cum,Sandick:2016zut,kowalska2017expectations}. Also, the thermal relic density of pure Bino DM tends to be (much) larger than the observed DM density; this is because, in the MSSM with a minimal flavor violation (MFV) structure, the Bino pair annihilation rate is suppressed and the DM decouples before its density is sufficiently depleted. Sizable left-right slepton mixing can play a role in enhancing the Bino annihilation rate, however a simultaneous match of the relic density and the $g_\mu-2$ anomaly is not possible unless one considers extra ingredients: In~\cite{Fukushima:2014yia} a CP-violating phase is introduced in the Bino-lepton-slepton couplings to drive an adhoc suppression of the contribution to the (CP-conserving) anomalous magnetic moment operator, while still allowing for a large Bino annihilation rate. As an alternative, we assume purely real Bino-lepton-slepton couplings and study the parameter space characterized by a small mass splitting between the Bino and the muon slepton driving the extra 1-loop contribution to $g_\mu-2$. For spectra sufficiently degenerate in mass, the muon slepton can delay the freeze out of the Bino through the so-called coannihilation effect, making the scenario cosmologically viable.

The model we study has very few parameters, essentially only 3 masses and one mixing angle, and constraints from $g_\mu-2$ and the DM relic density sharply cut through this parameter space. While recent studies (for example, see~\cite{Cox:2021nbo}) explore the case of DM production by Bino-smuon coannihilation in the context of addressing the muon $g-2$ anomaly within more typical realizations of the MSSM, we find that relaxing the assumption of MFV opens up a new region of parameter space in which the mass of the smuons can sit at the TeV scale. While there exists a viable parameter space in our model which is independent of the mechanism that provides for the chiral mixing of the smuons, we also consider the implications of the rather generic assumption that the off-diagonal element of the smuon mass matrix is associated with electroweak (EW) symmetry breaking in the SM. In particular, a trilinear coupling between the SM-like Higgs boson, left-handed smuon and right-handed smuon can both provide for the contribution to $g_\mu-2$ and can play an important role in coannihilation processes which deplete the Bino relic density. This scenario is clearly not natural from the point of view of fine-tuning, we will simply assume a Higgs sector which is SM-like and we will not address the issue of why small mass splittings occur in order for coannihilation effects to take place. On the other hand, we will discuss in detail theoretical self-consistency issues, such as perturbative unitarity and vacuum stability, illustrating trends which are relevant from a more general perspective as well as the model at hand. The scalar potentials of many BSM scenarios, including but not limited to that of the MSSM, can exhibit violations of perturbative unitarity or unstable EW vacua when the (dimensionful) couplings of trilinear scalar interactions become large. We explore the low-energy phenomenology of the model, focusing on the possibility of direct DM detection and commenting on indirect DM detection and LHC observables.

Our main results can be summarized by Figs.~\ref{fig:oh2iso1} and~\ref{fig:oh2iso2}, which can generally be characterized by two different regions of the parameter space. For Bino masses $\lesssim 400 \,$GeV, the parameter points which satisfy both $g_\mu-2$ and the DM relic density are similar to what has been typically found in previous scans of the MSSM, with the trilinear coupling only marginally impacting relic density. However, we again want to emphasize that the $g_\mu-2$ contribution in our model is typically subdominant in scans of MSSM parameter space. As the size of the trilinear coupling increases, we show that another region of parameter space opens up which can satisfy both $g_\mu-2$ and the DM relic density for Bino masses up to $\sim 1 \,$TeV.\footnote{Previous studies have demonstrated that satisfying the DM relic density is possible through coannihilations involving $\mathcal{O}({\rm TeV})$ scalars~\cite{Garny:2014waa,ElHedri:2017nny,Aboubrahim:2017aen,Davidson:2017gxx,Ellis:2018jyl,Abdughani:2018bhj,ElHedri:2018atj,Baker:2018uox,duan2019vacuum}.} In this parameter region extending out to higher Bino masses, we demonstrate how violations of perturbative unitarity manifest in the calculation of the smuon annihilation cross section and then perform a detailed analysis of both perturbative unitarity and EW vacuum stability in our model. Regarding signatures in low-energy phenomenology which could distinguish our model from other BSM scenarios which address $g_\mu-2$, we show that most parameter points are extremely challenging to probe using direct DM detection. Also, mass spectra around $\sim 1 \,$TeV with such small mass splittings between the Bino and lightest smuon are typically beyond the reach of searches for such particles at the LHC (for example, see Refs.~\cite{ATLAS:2019lff, ATLAS:2019lng,Dutta:2014jda,Han:2014aea,Dutta:2017nqv}). However, it has been shown that future lepton colliders with relatively large center of mass energies could be sensitive to these models~\cite{deBlas:2018mhx,Berggren:2013vna,Baum:2020gjj}.

The rest of the paper is organized as follows: In Sec.~\ref{sec:model} we describe the particle content and interactions in our model most relevant for the calculation of $g_\mu-2$ and the relic density, which are described in Sec.~\ref{sec:gm2} and Sec.~\ref{sec:rel}, respectively. We investigate constraints from perturbative unitarity and vacuum stability in Sec.~\ref{sec:pert}. We study the sensitivity of direct DM detection to our model in Sec.~\ref{sec:dd}. In Sec.~\ref{sec:con}, we conclude with a discussion summarizing our results and briefly comment on potentially interesting future work related to this model.

\section{Constructing the model} \label{sec:model}

The model contains a Bino $\tilde{B}^0$, with mass $M_{\tilde{B}}$. This is a spin 1/2 Majorana fermion, transforming as a singlet under the SM $SU(2)_L \times U(1)_Y$, $({\bf 1},0)$. We will assume throughout the paper that $\tilde{B}^0$ is the lightest BSM particle and stable, with the stability protected by a $\mathbb{Z}_2$ symmetry under which all BSM states introduced are odd. The Bino is coupled to the SM only through the muon and muon neutrino, via the terms:
\be
\mathcal{L} \supset - \lambda_{\tilde{\mu}_R} \,\tilde{\mu}_R^*  \bar{\tilde{B}}^0 P_R \mu
- \lambda_{\tilde{\mu}_L} \,\tilde{\mu}_L^*  \bar{\tilde{B}}^0 P_L \mu
- \lambda_{\tilde{\nu}} \, \tilde{\nu}_\mu^*  \bar{\tilde{B}}^0 \nu_\mu  + \text{h.c.},
\label{eq:Bmusmu}
\ee
where $P_R$ and $P_L$ are the right-handed and left-handed projectors, and we have introduced two electrically charged complex scalars, $\tilde{\mu}_R^*$ transforming as $({\bf 1},1)$ and the $SU(2)_L$ doublet $\tilde{l}_L = (\tilde{\nu}_\mu,\tilde{\mu}_L)^T$, which transforms as $({\bf 2},-1/2)$. While the different $\lambda$ couplings can in principle be arbitrary without significantly impacting the low-energy phenomenology of the model, we match them with those in the MSSM, namely:
\be
\lambda_{\tilde{\mu}_{R}} =  \sqrt{2}~g^\prime Y_{R} \quad \quad {\rm and} \quad \quad 
\lambda_{\tilde{\mu}_{L}} =  \lambda_{\tilde{\nu}} = \sqrt{2}~g^\prime Y_{L},
\label{eq:lambdaMSSM}
\ee
where $g^\prime$ is the SM hypercharge coupling. On the other hand, we consider a generic mixing for the two charged scalars starting from a fully general mass matrix,
\begin{eqnarray}
\mathcal{L} \supset -\left(\begin{matrix}
\tilde{\mu}_L^*&\tilde{\mu}_R^*
\end{matrix}\right)
\left(\begin{matrix}
m_{LL}^2&m_{LR}^2\\
m_{LR}^2&m_{RR}^2
\end{matrix}\right)\left(\begin{matrix}
\tilde{\mu}_L\\
\tilde{\mu}_R
\end{matrix}\right),
\end{eqnarray}
and diagonalizing it to find mass eigenstates we have
\be
\left(\begin{matrix}  \tilde \mu_1 \\  \tilde \mu_2 \end{matrix} \right)=
\left(\begin{matrix} \cos \theta_{\tilde{\mu}}  & -\sin \theta_{\tilde{\mu}}  \\ \sin \theta_{\tilde{\mu}}  & \cos \theta_{\tilde{\mu}}  
\end{matrix} \right)
\left(\begin{matrix}  \tilde \mu_L \\  \tilde \mu_R \end{matrix} \right)\,.
\ee
The convention we adopt is that $\tilde \mu_1$ is always lighter than $\tilde \mu_2$ and the mixing angle $\theta_{\tilde{\mu}}$ is in the interval $[-\pi/2,\pi/2)$. In the following, rather than using the entries of the mass matrix as free parameters, it is more convenient to refer to physical parameters, namely the two masses $M_{\tilde \mu_1}$, $M_{\tilde \mu_2}$, and $\theta_{\tilde{\mu}}$, or, equivalently to $M_{\tilde \mu_1}$, $\Delta M_{21}^2 \equiv M^2_{\tilde \mu_2} - M^2_{\tilde \mu_1}$ and $\theta_{\tilde{\mu}}$; the relative mapping is given by
\be
m_{LL}^2 = M_{\tilde \mu_1}^2 + [1 - \cos (2 \theta_{\tilde{\mu}} )]/2\cdot \Delta M_{21}^2, \quad
m_{RR}^2 = M_{\tilde \mu_1}^2 + [1 +\cos (2 \theta_{\tilde{\mu}} )]/2\cdot \Delta M_{21}^2, \quad
m^2_{LR} =  \sin (2 \theta_{\tilde{\mu}} )/2 \cdot\Delta M_{21}^2.
\ee
The muon sneutrino $\tilde{\nu}_\mu$, the left-handed neutral scalar we introduced above, does not appear in the BSM contribution to $g_\mu-2$, however it can play a role in the relic density computation; as we will show in Sec.~\ref{sec:rel}, we cannot simply assume it is very heavy and decouples. Inspired again by the MSSM, we write the muon sneutrino mass as
\be
M_{\tilde{\nu}_\mu}^2  \equiv M_{\tilde \mu_{1}}^2 + [1 - \cos (2 \theta_{\tilde{\mu}} )]/2\cdot \Delta M_{21}^2
- \Delta M^2_{W}\,.
\label{eq:snumass}
\ee
Under the assumption of minimal flavor violation in the MSSM one would simply have that $\Delta M^2_{W} = m_\mu^2-M_W^2 \cos 2\beta$, \ie, in the large $\tan \beta$ limit, $\Delta M^2_{W} \simeq M^2_{W}$ (here $m_\mu$ is the muon mass, $M_W$ the $W$ boson mass, and $\tan \beta$ the ratio between vacuum expectation values in the two Higgs doublet structure of the MSSM). In general, we will show that the parameter $\Delta M_{W}^2$ cannot be far from the weak scale, and, if $\Delta M_{W}^2>0$ as in the MSSM, there is a range of smuon masses and mixings for which $\tilde{\nu}_\mu$ is lighter than $\tilde \mu_1$:  In the limit $\theta_{\tilde{\mu}} \rightarrow 0$, when $\tilde \mu_1$ is almost purely left-handed and $M_{\tilde \mu_1} \simeq m_{LL}$, $\tilde{\nu}_\mu$ is the lightest slepton with $M_{\tilde{\nu}_\mu}^2 \simeq M_{\tilde \mu_1}^2 - \Delta M^2_{W}$; on the other hand, when the mixing angle increases, the mass ordering between $\tilde \mu_1$ and  $\tilde{\nu}_\mu$ can flip. In the opposite limit, when $|\theta_{\tilde{\mu}}| \rightarrow \pi/2$, $\tilde \mu_1$ is mostly right-handed, with $M_{\tilde \mu_1} \simeq m_{RR}$, while $\tilde \mu_2$ and $\tilde{\nu}_\mu$ can be (much) heavier.

The additional gauge invariant terms one can introduce involve couplings of the BSM scalars to the SM Higgs. Rather than considering a generic structure, our starting point will again be the MSSM. For simplicity, we consider the limit in which there is only one light Higgs (it would be $H_2^0$ in the MSSM jargon), which is SM-like and with its mass fine-tuned to the experimental value. This picture is equivalent to the ``decoupling limit" of the MSSM, in which the mass of the pseudoscalar $A$ is very heavy and the mixing angle between the two CP-even Higgs states, $\alpha$, is fixed such that $\sin(\beta-\alpha) \rightarrow 1$. We can then write the couplings for the trilinear terms involving the physical states that remain in the low energy theory, factoring out $g\, M_W$, where $g$ is the $SU(2)_L$ coupling,
\bea
y_{H_2^0 \tilde{\mu}_L \tilde{\mu}_L} &\rightarrow& \frac{1/2 -\sin^2 \theta_W}{\cos^2 \theta_W} \cos(2\beta)
\quad \quad  y_{H_2^0 \tilde{\mu}_L \tilde{\mu}_R} \rightarrow - \frac{\Delta M_{21}^2}{4\,M_W^2} \sin(2 \theta_{\tilde{\mu}}) \nonumber  \\
y_{H_2^0 \tilde{\mu}_R \tilde{\mu}_R} &\rightarrow & \frac{\sin^2 \theta_W}{\cos^2 \theta_W} \cos(2\beta)
\quad \quad \quad \quad \quad y_{H_2^0 \tilde{\nu}_\mu \tilde{\nu}_\mu} \rightarrow - \frac{1}{2\,\cos^2 \theta_W} \cos(2\beta)
\label{eq:higgscoup}
\eea
where $\theta_W$ is the Weinberg angle and we have neglected contributions $\propto m_\mu^2/M_W^2$ (in the following we will only consider the large $\tan \beta$ limit, with $\cos(2\beta)\rightarrow -1$). 

While we have taken a specific limit of the MSSM as a benchmark to define the trilinear couplings in our model, we again want to emphasize that the low-energy phenomenology of the model would remain qualitatively the same for different choices of couplings. As we have mentioned and discuss further in Sec.~\ref{sec:gm2}, the left-right smuon mixing is a key ingredient for a sizable BSM contribution to the anomalous magnetic moment. Over a large part of parameter space in the model outlined above, the particular origin of the left-right mixing does not significantly impact the calculation of $\Delta a_\mu$. Given the specific choices in Eq.~(\ref{eq:higgscoup}), the ``off diagonal" coupling which provides for the left-right smuon mixing, $y_{H_2^0 \tilde{\mu}_L \tilde{\mu}_R}$, only becomes relevant in the 1-loop diagrams which yield the dominant contribution to $g_\mu-2$ in the limit where this coupling becomes large. Even for $y_{H_2^0 \tilde{\mu}_L \tilde{\mu}_R} \gg 1$, we demonstrate that variations over several orders of magnitude can be compensated by $\mathcal{O}(1)$ changes in the Bino mass for fixed $\Delta a_\mu$. 

Similarly, in Sec.~\ref{sec:rel} we show that parts of the ``model-independent" parameter space which can satisfy $g_\mu-2$ can also satisfy the DM relic density largely independent from how the left-right smuon mixing is generated. We demonstrate how the calculation of the Bino relic density can be effected when the delicate cancellation between contributions to the cross sections for various processes involving SM gauge interactions is spoiled in scenarios with sizable smuon mixing angles and large mass splittings between the smuons. Placed within the context of our MSSM-like benchmark, large trilinear couplings can also directly enter into the calculation of cross sections for processes relevant to the Bino relic density. In either case, any associated changes to the cross sections can easily be absorbed into the Boltzmann suppression factors which are exponentially dependent on the mass splitting between the lightest sleptons and the Bino. The quartic scalar interactions, which we assume to take the form of the D-terms in the MSSM, have only a marginal effect on the relic density and a negligible role in the left-right mixing (at least for larger smuon mixing angles or heavier smuon masses). The specific form of the trilinear and quartic couplings are instead crucial when considering constraints from perturbative unitarity and vacuum stability; we thus give a more detailed description of the full scalar potential in Sec.~\ref{sec:pert}. 

\section{Constraints from the muon anomalous magnetic dipole moment}  \label{sec:gm2}

The leading extra contribution (the only contribution at 1-loop) to the muon anomalous magnetic dipole moment in our model is given by two diagrams. Each of these diagrams involves the Bino and one of the two smuons as virtual states running in the loop, with the external photon attached to the smuon. To lowest order in the muon mass, this contribution can be written as (see, \eg,~\cite{Moroi:1995yh})
\be
\Delta a_\mu \simeq \frac{{g^\prime}^2 Y_L Y_R}{16 \pi^2} \, \sin(2\,\theta_{\tilde{\mu}}) 
\frac{m_\mu}{M_{\tilde{B}}}  \left[L(r_1)- L(r_2)\right]\,,
\label{eq:deltamuth}
\ee
where the loop function is
\be
L(r) \equiv \frac{r}{(1-r)^2} \left[1+r+\frac{2\,r\ln r}{(1-r)}\right] \quad \quad {\rm and} \quad \quad r_i \equiv \frac{M_{\tilde{B}}^2}{M_{\tilde \mu_i}^2}\,.
\label{eq:deltamuths}
\ee 
A smoothly increasing function of $r$, $L(r)$ is 0 for $r=0$ and 1/3 for $r=1$. To get an idea for how $\Delta a_\mu$ depends on the parameters of our model, we focus on the limit in which the mass splitting between $M_{\tilde{B}}$ and $M_{\tilde \mu_1}$ is small, typically $\Delta_1 \equiv (M_{\tilde \mu_1}- M_{\tilde{B}})/M_{\tilde{B}} \lesssim 5-10\%$ for co-annihilation effects to sufficiently deplete the Bino relic density. If we further assume that $\Delta M_{21}^2$ is sizable compared to $M_{\tilde{B}}^2$, we can match the extra contribution to the muon anomalous magnetic dipole moment with the central value of $\Delta a_\mu^{exp}$ in Eq.~(\ref{eq:deltamuexp}),
\be \label{eq:DamuLargedm}
\frac{\Delta a_\mu}{25.1 \cdot 10^{-10}} \simeq \left(\frac{-\sin(2\,\theta_{\tilde{\mu}})}{2.6 \cdot 10^{-2}}\right) \left(\frac{100\;{\rm GeV}}{M_{\tilde{B}}}\right) \left(\frac{L_{0,2}+L_{1,2}\cdot (\Delta_1/0.1)+{\mathcal O} \left(\Delta_1^2 \right)}{0.23}\right) \, .
\ee
The loop function, $L$, in the expression above has been evaluated assuming a fixed ratio between the smuon masses, namely $M_{\tilde \mu_2} = 2\, M_{\tilde \mu_1}$, and expanded in $\Delta_1$, obtaining coefficients $L_{0,2}\simeq 0.19$ and $L_{1,2}\simeq 0.04$ (the picture is unchanged for another sample choice, \eg, if $M_{\tilde \mu_2} = 4\, M_{\tilde \mu_1}$, the associated are coefficients $L_{0,4}\simeq 0.28$ and $L_{1,4}\simeq 0.04$). 

For $M_{\tilde{B}} \sim 100 \,$GeV, matching $\Delta a_\mu^{exp}$ in Eq.~(\ref{eq:DamuLargedm}) requires either $\theta_{\tilde{\mu}}$ slightly smaller than 0 (when ${\tilde \mu_1}$ is mostly left-handed) or slightly larger than $-\pi/2$ (when ${\tilde \mu_1}$ is mostly right-handed). To match $\Delta a_\mu^{exp}$ for increasing $M_{\tilde{B}}$, $|\sin(2\,\theta_{\tilde{\mu}})|$ must also increase either along a ``left-handed branch" or a ``right-handed branch". This functional dependence in Eq.~(\ref{eq:DamuLargedm}) suggests that the two branches would join at an endpoint with maximal mixing and $\tilde{B}^0 \gtrsim 1 \,$TeV. However, as we discuss in Sec.~\ref{sec:rel}, $\Delta M_{21}^2 \gg M_W^2$ can lead to the violation of perturbative unitarity in the cross sections relevant for depleting the Bino relic density via coannihilation. Thus, if we keep $\Delta M_{21}^2$ of order $M_W^2$ instead and expand the expression for $\Delta a_\mu$ in the limit of large $M_{\tilde{B}}$, we find 
\be \label{eq:DamuLargemb}
\frac{\Delta a_\mu}{25.1 \cdot 10^{-10}} \simeq 0.090 \cdot \left(- \frac{\Delta M_{21}^2}{4 M_W^2} \sin(2\,\theta_{\tilde{\mu}}) \right)
\left(\frac{1\;{\rm TeV}}{M_{\tilde{B}}}\right) ^3
 \left(\frac{1+ 0.24 \cdot (\Delta_1/0.1)+ {\mathcal O} \left(\Delta_1^2 
 ,\Delta_{21}^2 \right)}{1.24}\right)
\ee
with $\Delta_{21}^2 \equiv \Delta M_{21}^2/M_{\tilde{B}}^2$. The above expression emphasizes the scaling of $\Delta a_\mu$ with the the parameter combination introduced in Eq.~(\ref{eq:higgscoup}) as the chirality flipping trilinear coupling, $y_{H_2^0 \tilde{\mu}_L \tilde{\mu}_R}$. For $M_{\tilde{B}}$ at the TeV scale and $\Delta M_{21}^2/(4 M_W^2) |\sin(2\,\theta_{\tilde{\mu}})| \sim 1$, the extra contribution to $g_\mu-2$ cannot match $\Delta a_\mu^{exp}$ due to the additional suppression $\propto \Delta M_{21}^2 / M_{\tilde{B}}^2$ relative to Eq.~(\ref{eq:DamuLargedm}). On the other hand, while the expansion in Eq.~(\ref{eq:DamuLargemb}) tends break down as $\Delta M_{21}^2$ becomes much larger than $M_W^2$ while keeping  $M_{\tilde{B}} \sim 1 \,$TeV, it suggests that $\Delta a_\mu$ can match the measured value for $\Delta M_{21}^2/(4 M_W^2) |\sin(2\,\theta_{\tilde{\mu}})| \sim \mathcal{O}(10)$. In Sec.~\ref{sec:pert}, we perform a detailed analysis of perturbative unitarity under the assumption that $y_{H_2^0 \tilde{\mu}_L \tilde{\mu}_R}$ provides for the chiral mixing of the smuons, in addition to an investigation of how large trilinear couplings can destabilize the EW vacuum.

\begin{figure}[t!]
\centering
\includegraphics[scale=0.54]{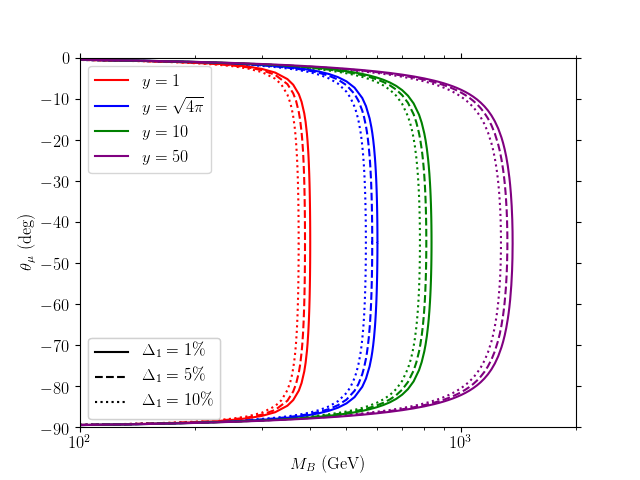}
\includegraphics[scale=0.54]{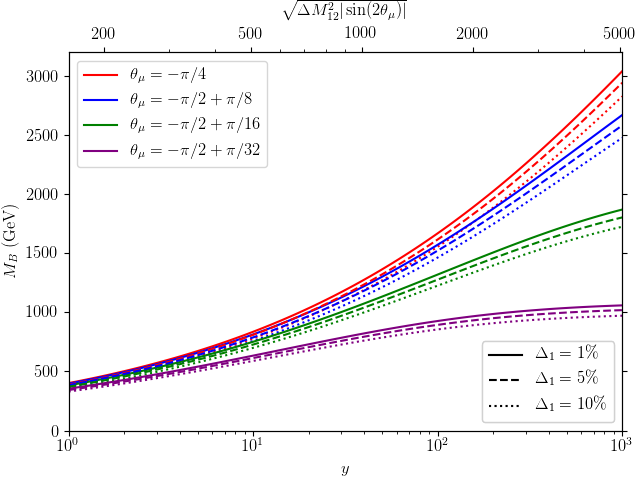}
\caption{{\sl Left panel:} Plot of the mixing angle $\theta_{\tilde{\mu}}$, in degrees, corresponding to an extra contribution to $g_\mu-2$ matching the central value in Eq.~(\ref{eq:deltamuexp}), for a given value of the Bino mass $M_{\tilde{B}}$, a few sample choices of the relative mass splitting between the Bino and the lightest smuon $\Delta_1$, and fixed values of the parameter $y$, which contains information on the mass of the heaviest smuon (see the definition in Eq.~(\ref{eq:y})). {\sl Right panel:} Plot of the Bino mass for which a $g_\mu-2$ match is possible versus the parameter $y$ and sample choices of $\theta_{\tilde{\mu}}$ and $\Delta_1$; labels at the top of the plot indicate the one-to-one correspondence between $y$ and the smuon mass squared differences $\Delta M_{21}^2$ for fixed $\theta_{\tilde{\mu}}$. As discussed in Sec.~\ref{sec:pert} under the assumption that $y$ is associated with a trilinear coupling between the SM Higgs and the smuons, models with values of $y \gg 1$ can be subject to constraints from perturbative unitarity and EW vacuum stability.}
\label{fig:thetamu}
\end{figure}

In Fig.~\ref{fig:thetamu} the extra contribution to the muon anomalous magnetic dipole moment, computed considering the full 1-loop result in Eq.~(\ref{eq:deltamuth}), matches $\Delta a_\mu^{exp}$ for all models displayed. In the left panel, values of the mixing angle $\theta_{\tilde{\mu}}$ are shown plotted against $M_{\tilde{B}}$, having fixed the relative mass splitting $\Delta_1$ between the lightest smuon and the Bino. The mass of the heaviest smuon is fixed by selecting a given value for the parameter
\be
y \equiv | y_{H_2^0 \tilde{\mu}_L \tilde{\mu}_R} | = \Delta M_{21}^2/(4 M_W^2) \cdot |\sin(2\,\theta_{\tilde{\mu}})|\,.
\label{eq:y}
\ee
As sketched above, at low $M_{\tilde{B}}$ one can see the left-handed and the right-handed branches, while $y$ sets the maximum Bino mass scale at which for they join. Larger $y$ corresponds to a heavier maximum $M_{\tilde{B}}$, with a more mild dependence on the precise value of $\Delta_1$. For $y=1$ we see that $M_{\tilde{B}} \lesssim 400 \,$GeV and for $y=10$ we have $M_{\tilde{B}} \lesssim 800 \,$GeV, with the latter maximum Bino mass somewhat smaller than what is suggested by the approximation in Eq.~(\ref{eq:DamuLargemb}) after accounting for the full 1-loop result for  $\Delta a_\mu$.

However, in the case which fixes $y=50$ in the left panel of Fig.~\ref{fig:thetamu}, we see that satisfying $g_\mu-2$ with $M_{\tilde{B}} \sim 1 \,$TeV is indeed possible for large enough values of $y$. In the right panel, $M_{\tilde{B}}$ is plotted against $y$ for fixed values of $\theta_{\tilde{\mu}}$ and $\Delta_1$. We see that models satisfying $\Delta a_\mu^{exp}$ with $M_{\tilde{B}} \gtrsim 1 \,$TeV require larger values of $y \gtrsim 20$, and hence larger values of the mass splitting between heavy and light smuons (at given $\theta_{\tilde{\mu}}$ there is of course a one-to-one match between $y$ and $\Delta M_{21}^2$, as indicated along the top of the plot). Going to even larger Bino masses while still matching $\Delta a_\mu^{exp}$ requires much larger $y$ and smuon mixing angles closer to maximal. For example, in the case of maximal mixing $\theta_{\tilde{\mu}} = -\pi/4$, a Bino can be as heavy as about 2 TeV only for $y \simeq 190$, corresponding to $\sqrt{\Delta M_{21}^2} \simeq 2.8$~TeV.

While the numerical results reported in this Section hold for the specific choice of $\lambda_{\tilde{\mu}_{R}}$ and $\lambda_{\tilde{\mu}_{L}}$ in Eq.~(\ref{eq:lambdaMSSM}), the general picture is unchanged for generic Bino-smuon-muon couplings. The replacement ${g^\prime}^2 Y_L Y_R \rightarrow \lambda_{\tilde{\mu}_{L}} \lambda_{\tilde{\mu}_{R}}/2$ in the expression for $\Delta a_\mu$ given by Eq.~(\ref{eq:deltamuth}) would imply different ``plateau values" of $\theta_{\tilde{\mu}}$ at small Bino masses in Fig.~\ref{fig:thetamu} (with, eventually, a flip in the sign of $\theta_{\tilde{\mu}}$). For large $M_{\tilde{B}}$, generic Bino-smuon-muon couplings could be absorbed into a different definition of the parameter $y$, for instance $y^\prime \propto \lambda_{\tilde{\mu}_{R}} \lambda_{\tilde{\mu}_{L}} \Delta M_{21}^2 \sin(2\,\theta_{\tilde{\mu}})$.
 
\section{Constraints from the relic density calculation}  \label{sec:rel}

A stable massive particle, with weak interaction couplings to the thermal bath of the early Universe, tends to have a relic density the order of the measured DM density of the Universe. This is the celebrated ``WIMP miracle," often summarized with a formula stating an approximate inverse relation between the relic abundance and the thermally averaged pair annihilation cross section for the DM particle computed at the freeze-out temperature $T_f$,
\be
\Omega h^2 \sim 0.1 \left(\frac{1 \; {\rm pb}}{\langle \sigma v \rangle(T_f)}\right)\,.
\label{eq:roft}
\ee
This approximation is best applied to the case of ``vanilla" thermal relics, in which the pair annihilation is not $s$-wave suppressed: Since $T_f$ is generically about 5\% of the DM particle mass $M$, one can consider the expansion
\be
\langle \sigma v \rangle(T_f) = \sigma_0 +  \sigma_1 \cdot \left(\frac{T_f}{M} \right) +  {\mathcal O} \left(\frac{T_f^2}{M^2}\right)\,,
\ee
to highlight that $s$-wave annihilations provide a contribution to all terms, $\sigma_i$, in the series and the first non-zero coefficient from higher wave contributions to the thermally averaged cross section arises from $p$-wave annihilations in $\sigma_1$. In the MSSM with a minimal flavor violation structure, Bino annihilation into light SM fermions is $s$-wave suppressed. This is because the annihilation, at zero orbital angular momentum, of a pair of identical Majorana fermions (total spin equal to 0) into a pair of chiral fermions requires a chirality flip in order to conserve total angular momentum; if the flip can proceed only via an insertion of the final state fermion mass, then $\sigma_0 \propto m_f^2/M^2$,  which is sharply suppressed for $m_f \ll M$ (\eg~for the annihilation of Bino DM in the so-called ``bulk region" of the MSSM). 

For the model we are considering, the picture is in principle different: At tree level, Binos can annihilate only into $\nu_\mu \bar{\nu}_\mu$ and $\mu^- \mu^+$. Bino annihilation into muons arises from the first two interaction terms of the Lagrangian in Eq.~(\ref{eq:Bmusmu}). The process is mediated by smuons in the $t-$ and $u-$channels, and---most relevantly---the chirality flip can be provided for by the explicit left-right mixing introduced in the smuon mass matrix. This is the same mechanism which allows for a sizable contribution to $\Delta a_\mu$ and, thus, Bino annihilation into muons can be correlated with $g_\mu-2$ in our model. To lowest order in the muon mass, the $s$-wave contribution to the cross section for Bino annihilation into muons is given by (see also, \eg,~\cite{Fukushima:2014yia})
\be
 \sigma_{0,\tilde{B}^0\tilde{B}^0\rightarrow \mu^- \mu^+} \simeq  \frac{{g^\prime}^4 Y_L^2 Y_R^2}{8 \pi} \, \sin^2(2\,\theta_{\tilde{\mu}}) \frac{1}{M_{\tilde{B}}^2}  \left(\frac{1}{1+r_1}-\frac{1}{1+r_2}\right)^2 \,.
\ee
Matching this expression to $\Delta a_\mu$ from Eq.~(\ref{eq:deltamuth}), the cross section can be rewritten as
\bea
 \sigma_{0,\tilde{B}^0\tilde{B}^0\rightarrow \mu^- \mu^+}  &\simeq&  \frac{32 \pi^3 (\Delta a_\mu)^2}{m_\mu^2} 
 \left[\frac{(1+r_1)^{-1}-(1+r_2)^{-1}}{L(r_1)- L(r_2)}\right]^{2}  \nonumber \\
 &\simeq& 2.2 \cdot10^{-4} \; {\rm pb} \;
  \left[\frac{(1+r_1)^{-1}-(1+r_2)^{-1}}{L(r_1)- L(r_2)}\right]^{2} \;
 \left(\frac{\Delta a_\mu}{25.1 \cdot 10^{-10}}\right)^2.
\eea
Note the function in the square brackets in the expression above is at most $\sqrt{8/3}$ for any $1\ge r_1 \ge r_2$. Therefore, at any point in the parameter space of our model for which $\Delta a_\mu \simeq \Delta a_\mu^{exp}$, $\sigma_0$ is much smaller than the annihilation cross section needed to satisfy the rule-of-thumb in Eq.~(\ref{eq:roft}). As a result, $s$-wave pair annihilation cannot be the mechanism providing for the thermal relic density that matches the DM density in the Universe. 

As a related issue, it follows that the prospect for indirect DM detection of annihilation signals from DM halos (in which DM particles have typically very small velocities) are not encouraging in our scenario. One notable exception may be for annihilation signals from extremely overdense DM environments, such as the ``DM spike" which could form around a black hole after its adiabatic growth~\cite{Gondolo:1999ef,Ullio:2001fb,Bertone:2005xz}; we are not going to discuss these scenarios further here. Notice that the correlation between Bino $s$-wave annihilation and $\Delta a_\mu$, as well as the associated difficulty of simultaneously satisfying the relic DM abundance and $g_\mu-2$, holds for any generic real Bino-smuon-muon couplings $\lambda_{\tilde{\mu}_{R}}$ and $\lambda_{\tilde{\mu}_{L}}$. The tension between the relic density arising from $s$-wave DM annihilation and $\Delta a_\mu$ can only be softened by introducing CP-violating phases, as suggested in~\cite{Fukushima:2014yia}. 

The leading $p$-wave contribution to the cross section for Bino annihilation into muons from chirality conserving processes can be approximated by
\be
  \sigma_{1,\tilde{B}^0\tilde{B}^0\rightarrow \mu^- \mu^+} \cdot \left(\frac{T_f}{M_{\tilde{B}}} \right)
  \sim 0.5  \; {\rm pb} \left(\frac{100  \; {\rm GeV}}{M_{\tilde \mu_1}} \right)^2\,,
\ee
and becomes relevant only if the particle spectrum is rather light. However, such light spectra are in the parameter region where smuon masses have been excluded by collider searches in a model independent way~\cite{ATLAS:2019lng,ATLAS:2019lff}. For generic choices of $\lambda_{\tilde{\mu}_{R}}$ and $\lambda_{\tilde{\mu}_{L}}$, the tension betwen satisfying the relic density and constraints from colliders can be relaxed to a certain extent since the $p$-wave contribution above scales with the fourth power of one of these couplings and is not necessarily correlated with $\Delta a_\mu$ or the smuon production cross section relevant for a collider search. We are not going to follow this route, nor the suggestion to consider CP violation as in~\cite{Fukushima:2014yia}, but rather concentrate on our minimal setup and explore the consequences of compressed particle spectra.  


\begin{table}[t!]
\centering

\begin{tabular}{p{5cm}p{1.5cm}p{1.5cm}p{1.5cm}p{0.5cm}}
 & \multicolumn{4}{c}{Diagrams} \\ \cline{2-5}
Process & s  & t & u & p \\
\hline
~ & \\[-2.5ex]
$\tilde{B}^0 \tilde{B}^0 \into \nu_\mu  \bar{\nu}_\mu$ & &  $\tilde{\nu}_\mu$ & $\tilde{\nu}_\mu$ \\
$\tilde{B}^0 \tilde{B}^0 \into \mu^- \mu^+$ &  & $\tilde{\mu}_{1,2}$ & $\tilde{\mu}_{1,2}$ \\[1ex]
\hline
~ & \\[-2.5ex]
$\tilde{\mu}_i \tilde{B}^0  \into Z^0  \mu^-,\,\gamma  \mu^-,\,H^0  \mu^-$   & $\mu^-$   & $\tilde{\mu}_{1,2}$   \\
$\tilde{\mu}_i \tilde{B}^0  \into W^-  \nu_\mu $   & $\mu^-$  & & $\tilde{\nu}_\mu$ &  \\[1ex]
\hline
~ & \\[-2.5ex]
$\tilde{\mu}_i \tilde{\mu}_j^* \into f \bar{f}$  & $H^0,Z^0,\gamma$ & \\
$\tilde{\mu}_i \tilde{\mu}_j^* \into \mu^- \mu^+$  & $H^0,Z^0,\gamma$ & $\tilde{B}^0$ \\
$\tilde{\mu}_i \tilde{\mu}_j \into \mu^- \mu^-$  & & $\tilde{B}^0$ \\
$\tilde{\mu}_i \tilde{\mu}_j^* \into W^- W^+$  & $H^0,Z^0,\gamma$ & $\tilde{\nu}_\mu$  & & p \\
$\tilde{\mu}_i \tilde{\mu}_j^* \into Z^0 Z^0,\,H^0 H^0$  & $H^0$ & $\tilde{\mu}_{1,2}$  & $\tilde{\mu}_{1,2}$ & p \\
$\tilde{\mu}_i \tilde{\mu}_j^* \into Z^0 \gamma$  &  & $\tilde{\mu}_{1,2}$  & $\tilde{\mu}_{1,2}$ & p \\
$\tilde{\mu}_i \tilde{\mu}_j^* \into \gamma \gamma,\,\gamma H^0$  &  & $\tilde{\mu}_{1,2}$  & $\tilde{\mu}_{1,2}$ & \\
$\tilde{\mu}_i \tilde{\mu}_j^* \into Z^0 H^0$  & $Z^0$ & $\tilde{\mu}_{1,2}$  & $\tilde{\mu}_{1,2}$ & \\[1ex]
\hline
~ & \\[-2.5ex]
$\tilde{\nu}_\mu \tilde{B}^0  \into Z^0  \nu_\mu,\,H^0 \nu_\mu$   & $ \nu_\mu$   & $\tilde{\nu}_\mu$   \\
$\tilde{\nu}_\mu \tilde{B}^0  \into W^+  \mu^-$   & $ \nu_\mu$  &  & $\tilde{\mu}_{1,2}$   \\[1ex]
\hline
~ & \\[-2.5ex]
$\tilde{\nu}_\mu \tilde{\nu}_\mu^* \into f \bar{f}$  & $H^0,Z^0$ & \\
$\tilde{\nu}_\mu \tilde{\nu}_\mu^* \into \nu_\mu \bar{\nu}_\mu$  & $Z^0$ & $\tilde{B}^0$ \\
$\tilde{\nu}_\mu \tilde{\nu}_\mu^* \into \nu_\mu \nu_\mu$  & & $\tilde{B}^0$ \\
$\tilde{\nu}_\mu \tilde{\nu}_\mu^* \into W^- W^+$  & $H^0,Z^0$ & $\tilde{\mu}_{1,2}$  & & p \\
$\tilde{\nu}_\mu \tilde{\nu}_\mu^* \into Z^0 Z^0,\,H^0 H^0$  & $H^0$ & $\tilde{\nu}_\mu$  & $\tilde{\nu}_\mu$ & p \\
$\tilde{\nu}_\mu \tilde{\nu}_\mu^* \into Z^0 H^0$  & $Z^0$ & $\tilde{\nu}_\mu$  & $\tilde{\nu}_\mu$ & \\[1ex]
\hline
~ & \\[-2.5ex]
$\tilde{\nu}_\mu \tilde{\mu}_i^* \into f_u \bar{f}_d$  & $W^+$  \\
$\tilde{\nu}_\mu \tilde{\mu}_i^* \into \nu_\mu \mu^+$  & $W^+$ & $ \tilde{B}^0$ \\
$\tilde{\nu}_\mu \tilde{\mu}_i \into \nu_\mu \mu^-$  & & $ \tilde{B}^0$ \\
$\tilde{\nu}_\mu \tilde{\mu}_i^* \into W^+ Z^0$  & $W^+$ & $\tilde{\mu}_{1,2}$ & $\tilde{\nu}_\mu$ & p \\
$\tilde{\nu}_\mu \tilde{\mu}_i^* \into W^+ \gamma$  & $W^+$ & $\tilde{\mu}_{1,2}$ &  & p \\
$\tilde{\nu}_\mu \tilde{\mu}_i^* \into W^+ H^0$  & $W^+$ & $\tilde{\mu}_{1,2}$ & $\tilde{\nu}_\mu$ \\[1ex]
\hline
\end{tabular}
\caption{Included coannihilation processes through $s-$, $t-$, $u-$channels 
and four-point interactions (p).}
\label{tab:coanns}
\end{table}

Consider a setup in which the DM candidate is the lightest particle among a set of BSM states that share a quantum number and are all in thermal equilibrium in the early Universe. The DM candidate is, thus, stable and states with mass splittings relative to the DM particle no larger than about $T_f$ have abundances at freeze out comparable to the DM state. Moreover if the states nearly degenerate in mass with the DM couple to the SM heat bath significantly more strongly than the DM particle, the slightly heavier states would keep the ensemble of BSM states in equilibrium for a longer time and further deplete the DM density before freeze-out. This effect is typically dubbed ``coannihilation"~\cite{Binetruy:1983jf,Griest:1990kh} and the particles involved are usually referenced as ``coannihilating" particles. Coannihilation can be described by a set of coupled Boltzmann equations. Since heavier states are expected to decay into the lightest stable species shortly after decoupling and one is usually interested only in the final DM density, it is possible to reformulate the problem in terms of a single density evolution equation~\cite{Griest:1990kh,Edsjo:1997bg,Edsjo:2003us},
\be
  \frac{dn}{dt} =
  -3Hn - \langle \sigma_{\rm{eff}} v \rangle(T) 
  \left( n^2 - n_{\rm{eq}}^2 \right) \, .
   \label{eq:Boltzmann}
\ee
In Eq. (\ref{eq:Boltzmann}), $n=\sum_i n_i$ is the sum of the number densities of all coannihilating particles, $n_{\rm{eq}}$ the analogous quantity for thermal equilibrium distributions, $H$ the Hubble parameter, and $\langle \sigma_{\rm eff} v \rangle$ an effective thermally-averaged annihilation cross section. The latter is a sum of thermally averaged annihilation cross sections for any pair of coannihilating states, $i$ and $j$, weighted over equilibrium densities,
\be
  \langle \sigma_{\rm{eff}}v \rangle(T) = \sum_{ij} \langle \sigma_{ij}v \rangle(T)  \frac{n_i^{eq}(T) n_j^{eq}(T)}{[n^{eq}(T)]^2} =
  \frac{\int_0^\infty
  dp_{\rm{eff}} p_{\rm{eff}}^2 W_{\rm{eff}}(s) \,K_1 \!\left(
  \frac{\sqrt{s}}{T} \right) } { m_1^4 T \left[ \sum_i \frac{g_i}{g_1}
  \frac{m_i^2}{m_1^2} \,K_2 \!\left(\frac{m_i}{T}\right) \right]^2}\,.
  \label{eq:sigmavefffin2}
\ee
In the second equivalence above, $K_{h}(x)$, ($h=1,2$) are the modified Bessel functions of the second kind of order $h$, $m_i$ and $g_i$ the mass and number of internal degrees of freedom (statistical weights) for the particle $i$ ($i=1$ labels the lightest state),  $p_{\rm{eff}}$ an effective momentum defined through the usual Mandelstam variable $s$ as $s= 4p_{\rm{eff}}^2 + 4m_1^2$, and $W_{\rm eff}$ the effective annihilation rate given by
\be
  W_{\rm{eff}} (s) = \sum_{ij}\frac{p_{ij}}{p_{11}}
  \frac{g_ig_j}{g_1^2} W_{ij} = 
  \sum_{ij} \sqrt{\frac{[s-(m_{i}-m_{j})^2][s-(m_{i}+m_{j})^2]}
  {s(s-4m_1^2)}} \frac{g_ig_j}{g_1^2} W_{ij}.
  \label{eq:weff}
\ee
For the coannihilation of particles $i$ and $j$, 
$W_{ij}$ is the annihilation rate per unit volume and unit time,
\be 
  W_{ij} = 4 p_{ij} \sqrt{s} \sigma_{ij} = 4 \sigma_{ij} \sqrt{(p_i
  \cdot p_j)^2 - m_i^2 m_j^2} = 4 E_{i} E_{j} \sigma_{ij} v_{ij}\,,
  \label{eq:Wijcross}
\ee
where
\be
  p_{ij} =
  \left[s-(m_i+m_j)^2\right]^{1/2} \cdot \left[s-(m_i-m_j)^2\right]^{1/2} /(2\sqrt{s})
\ee
is the common magnitude of the 3-momentum of particles $i$ and $j$ in the center-of-mass frame of the $i$-$j$ pair.

Returning to the specific model we consider in this study and the computation of the Bino relic density, processes involving states besides the Bino can indeed contribute to the effective annihilation rate. Both the smuons, introduced as key ingredient for satisfying $g_\mu-2$, and the sneutrino, introduced for theoretical consistency, may have rates for pair annihilations and coannihilations with Binos larger than the Bino pair annihilation rate. If one or more of these scalars are sufficiently degenerate in mass with the Bino then its relic density can be depleted to the level favoured by cosmological measurements. Table~\ref{tab:coanns} contains the full list of annihilation and coannihilation processes which are included in our analysis; the relic density computation is then performed via a proper implementation of the model in the DarkSUSY package~\cite{bringmann2018darksusy}.

\begin{figure}[t!]
\centering
\includegraphics[scale=0.54]{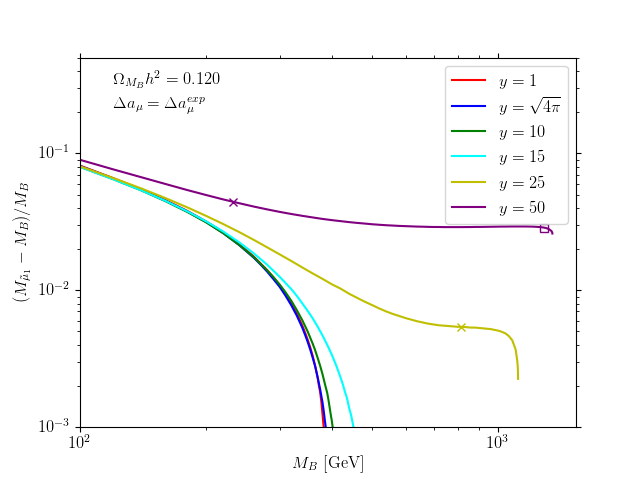}\includegraphics[scale=0.54]{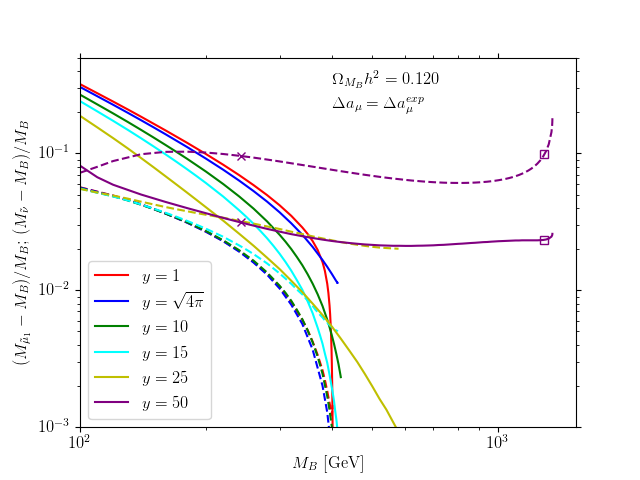}
\caption{Relative mass spitting between lightest smuon and Bino (solid lines) and sneutrino and Bino (dashed lines) required for coannihilation processes to drive the thermal relic density of the Bino to match the observed dark matter density. A few values of the smuon mass splitting parameter $y$ have been selected along the right-handed branch (left panel) and the left-handed branch (right panel). All models displayed match the $g_\mu-2$ excess. As discussed in Sec.~\ref{sec:pert} for cases with larger $y$ associated with a trilinear coupling between the SM Higgs and the smuons, we also indicate constraints for $M_{\tilde{B}}$ along the respective curves arising from perturbative unitarity (square) and EW vacuum stability (`x').}
\label{fig:oh2iso1}
\end{figure}

In Fig.~\ref{fig:oh2iso1}, for fixed values of the parameter $y$ and models matching the central value in $\Delta a_\mu^{exp}$, we show the mass splitting between the Bino and coannihilating states which yields a Bino thermal relic density matching the central value of the dark matter density measured by Planck~\cite{Planck2018values},
\be
  \Omega_{DM} h^2 = 0.11933 \pm 0.00091.
\ee 
The left panel refers to models on the ``right-handed" branch (RHB), while the right panel to the ``left-handed" one (LHB); with the exception of the case in which $y=50$, for the sample set of $y$ displayed the two branches do not join. On the RHB the relevant quantity is the mass splitting between the Bino and the (mostly right-handed) lightest smuon. For the LHB, we have fixed the mass splitting parameter $\Delta M^2_{W}$ in the sneutrino mass Eq.~(\ref{eq:snumass}) to its MSSM value, $\Delta M^2_{W} \simeq M^2_{W}$. From the related discussion of the mass spectrum in Sec.~\ref{sec:model}, recall that small smuon mixing angles imply $M_{\tilde{\nu}_\mu}^2 \lesssim M_{\tilde \mu_1}^2$ and the mass hierarchy can flip as the mixing angle increases. As shown in the left panel of Fig.~\ref{fig:thetamu}, satisfying $g_\mu-2$ requires the smuon mixing angle to become larger as $M_{\tilde{B}}$ increases. Also, the increase in the mixing angle must be more pronounced at smaller $M_{\tilde{B}}$ for smaller values of $y$. Thus, at small Bino masses and low $y$ on the LHB, the sneutrino is the next-to-lightest BSM state and its coannihilations drive the relic density. At moderate values of $y$, the lightest smuon (in this case mostly left-handed) may become lighter than the sneutrino for models that satisfy both the relic density and $g_\mu-2$. For example, the lightest smuon becomes lighter than the sneutrino at Bino masses larger than about 350~GeV for $y=15$, and larger than about 220~GeV for $y=25$. 

Another point worth noting: At values of $y \lesssim 15$ the parameter dependence of models that satisfy the relic density follows from the intuition that increases to the Bino mass must be compensated for by smaller mass splittings between the Bino and the coannihilating scalars. As the Bino mass increases, the masses of the scalars increase as well and the rates of the associated annihilation and coannihilation processes are suppressed. A decrease in the mass splitting can increase the weights of these processes in the effective thermally-averaged annihilation cross section. The larger the Bino mass, the smaller the mass splitting, until the effect saturates at a maximum mass and zero mass splitting. Thus, incorporating the relic density constraint sets an upper bound $M_{\tilde{B}} \lesssim 400 \,$GeV virtually independent of $y \lesssim 15$, which only enters marginally in the setting the effective annihilation rate. With the exception of the case with the smallest $y=1$, this upper bound on the Bino mass is more stringent than those which arise from requiring $g_\mu-2$ alone. 

Going to larger values of $y$ in Fig.~\ref{fig:oh2iso1}, we see the trend can change drastically. Specifically, for $y=25$ (RHB) and $y=50$ (RHB and LHB), we see that satisfying the relic density and $g_\mu-2$ for larger Bino masses requires the mass splittings to remain roughly constant or even become larger. In contrast to the cases with $y \lesssim 15$, this trend suggests that the rates for the most relevant annihilation and coannihilation processes can grow for larger particle masses and a corresponding increase in the mass splitting must decrease the weights of these processes in the effective thermally-averaged annihilation cross section. We investigate the manifestation of this peculiar behavior in the cross sections most relevant for the calculation of the relic density at large $y$ at the end of this Section. Also, assuming that $y$ is associated with a trilinear coupling between the SM Higgs and the smuons, we perform a detailed analysis of perturbative unitarity and EW vacuum stability in Sec.~\ref{sec:pert}. We indicate the results of the analysis in Sec.~\ref{sec:pert} with the colored markers along curves for larger $y$ in Fig.~\ref{fig:oh2iso1}. For $M_{\tilde{B}}$ larger than the `x' along a given curve, such models have shortlived metastable EW vacua and, for $M_{\tilde{B}}$ larger than the square along a given curve, such models are constrained by perturbative unitarity. We see that these conditions arising from theoretical self-consistency can severely restrict the parameter space of our model. 

\begin{figure}[t!]
\centering
\includegraphics[scale=0.54]{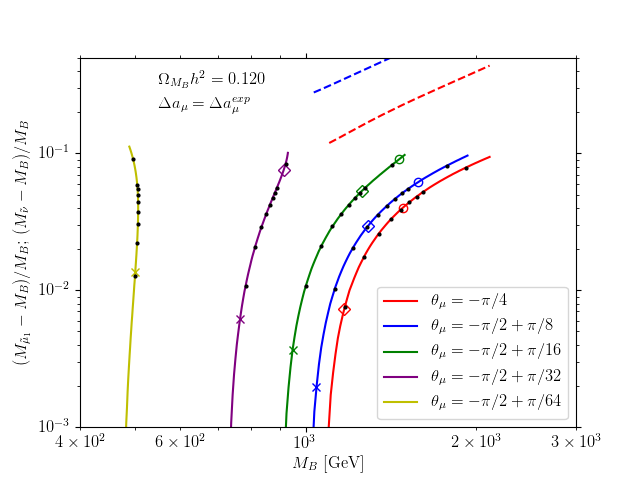}
\includegraphics[scale=0.54]{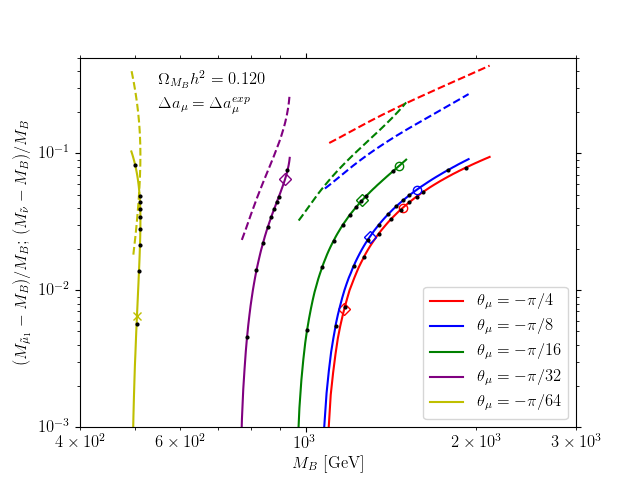}
\caption{The same as in Fig.~\ref{fig:oh2iso1}, but choosing a few sample values of the mixing angle $\theta_{\tilde{\mu}}$ and varying $y$ in the range [1,300]. Along each line, the small black markers indicate values of $y$, starting from the smallest mass splitting, $y=30$, 40, 50, 60, 70, 80, 90, 100, 200 and the endpoints at $y=300$; for $\theta_{\tilde{\mu}} = -\pi/2+\pi/64$ only, the black marker at the smallest mass splitting indicates $y=20$. Larger colored markers correspond to upper limits on $y$ along the curves from perturbative unitarity (square, circle) and EW vacuum stability (`x'), see Sec.~\ref{sec:pert} for details. Note that EW vacuum stability constrains all models with $\theta_{\tilde{\mu}} = -\pi/4$ displayed in the left panel and all models in the right panel are constrained except for a subset of those with $\theta_{\tilde{\mu}} = - \pi / 64$. Perturbative unitarity constrains no models displayed for either $\theta_{\tilde{\mu}} = - \pi /2 +  \pi / 64$ or $\theta_{\tilde{\mu}} = - \pi / 64$.}
\label{fig:oh2iso2}
\end{figure}

Before considering the theoretical self-consistency of our model in detail, we further explore the parameter space at large $y$ to identify regions which can satisfy both $g_\mu-2$ and the relic density for $M_{\tilde{B}} \gtrsim 400 \,$GeV. This parameter space is best illustrated in scans with a fixed mixing angle $\theta_{\tilde{\mu}}$ and varying $y$, as shown in Fig.~\ref{fig:oh2iso2}. For mixing angles fixed to $\theta_{\tilde{\mu}} = - \pi /2 +  \pi / 64$ (RHB) and $\theta_{\tilde{\mu}} = - \pi / 64$ (LHB), $M_{\tilde{B}} \simeq 500 \,$GeV is determined by $\Delta a_\mu$ almost independently of $y$. Also, since $y$ is large enough in these cases such that coannihilation processes drive the relic density, increases in $y$ enhance the effective annihilation rate and must be compensated for by larger mass splittings. For larger smuon mixing angles, the relationship between the mass splittings and $y$ is similar but the $M_{\tilde{B}}$ required to satisfy $g_\mu-2$ for a given value of $y$ largely follows from the parameter dependence of $\Delta a_\mu$ shown in the right panel of Fig.~\ref{fig:thetamu}. 

For the higher mass scales associated with the coannihilating particles in models with larger mixing angles shown in Fig.~\ref{fig:oh2iso2}, the contributions to the most relevant cross sections from terms involving the trilinear coupling are suppressed. However, as for the cases with large $y$ in Fig.~\ref{fig:oh2iso1}, the cross sections for processes which involve gauge interactions can grow with the coannihilating particle masses in models with moderately large $y$ and sizable left-right mixing. Again, a corresponding increase in the mass splitting is necessary to compensate for this peculiar effect. Similarly to Fig.~\ref{fig:oh2iso1}, the colored markers along the curves in Fig.~\ref{fig:oh2iso2} indicate the largest $y$ value along a given curve which is consistent with a sufficiently long-lived metastable EW vacuum and perturbative unitarity. We can see that vacuum stability constrains all points show which assume maximal smuon mixing while severely restricting the viable $y$ for other mixing angles. However, for $\theta_{\tilde{\mu}} = - \pi /2 +  \pi / 8$, points with $M_{\tilde{B}} \simeq 1 \,$TeV remain which can satisfy both $g_\mu-2$ and the relic density.

For cases considered above with arbitrarily large $y$, it is clear that that the relic density constraint is not setting an upper limit on the Bino mass because of the peculiar behavior of the cross sections which yield the dominant contributions to the effective annihilation rate. To illustrate this point further, we take one step back and we rewrite the effective thermally-averaged annihilation cross section from Eq.~(\ref{eq:sigmavefffin2}) as
\begin{equation} 
  \langle \sigma_{\rm{eff}}v \rangle = \int_0^\infty
  dp_{\rm{eff}} \frac{W_{\rm{eff}}(p_{\rm{eff}})}{4\,E_{\rm{eff}}^2} 
  \kappa(p_{\rm{eff}},T)\;,
  \label{eq:sigmavefffin3}
\end{equation}
where $E_{\rm{eff}} = (p_{\rm{eff}}^2 + M_{\tilde{B}}^2)^{1/2}$ is the energy per particle in the center of mass frame for Bino pair annihilation. The term we have isolated, $W_{\rm{eff}}/4{E^2_{\rm{eff}}}$, can be thought of as an effective $\sigma v$ term (compare with Eq.~(\ref{eq:Wijcross})). In the $p_{\rm{eff}}\rightarrow 0$ limit, it reduces to the DM annihilation rate at zero temperature, which is the relevant quantity for indirect DM detection of signals from DM pair annihilation. With Eq.~(\ref{eq:sigmavefffin3}) in this form, the function $\kappa$ contains the Boltzmann factors and the phase-space integrand term from Eq.~(\ref{eq:sigmavefffin2}). In effect, $\kappa$ can be interpreted as a window function that, at a given temperature $T$, selects the range of $p_{\rm{eff}}$ which is relevant in the thermal average. The phase-space integrand term dominates at small $p_{\rm{eff}}$ such that $\kappa=0$ at $p_{\rm{eff}}=0$. The function $\kappa$ exhibits a peak at intermediate $p_{\rm{eff}}$ and then rapidly decreases with larger $p_{\rm{eff}}$ due to the Boltzmann suppression in the thermal particle distributions; the position and height of the peak depends on the temperature considered and on the particles involved.

\begin{figure}[t!]
\centering
\includegraphics[scale=0.54]{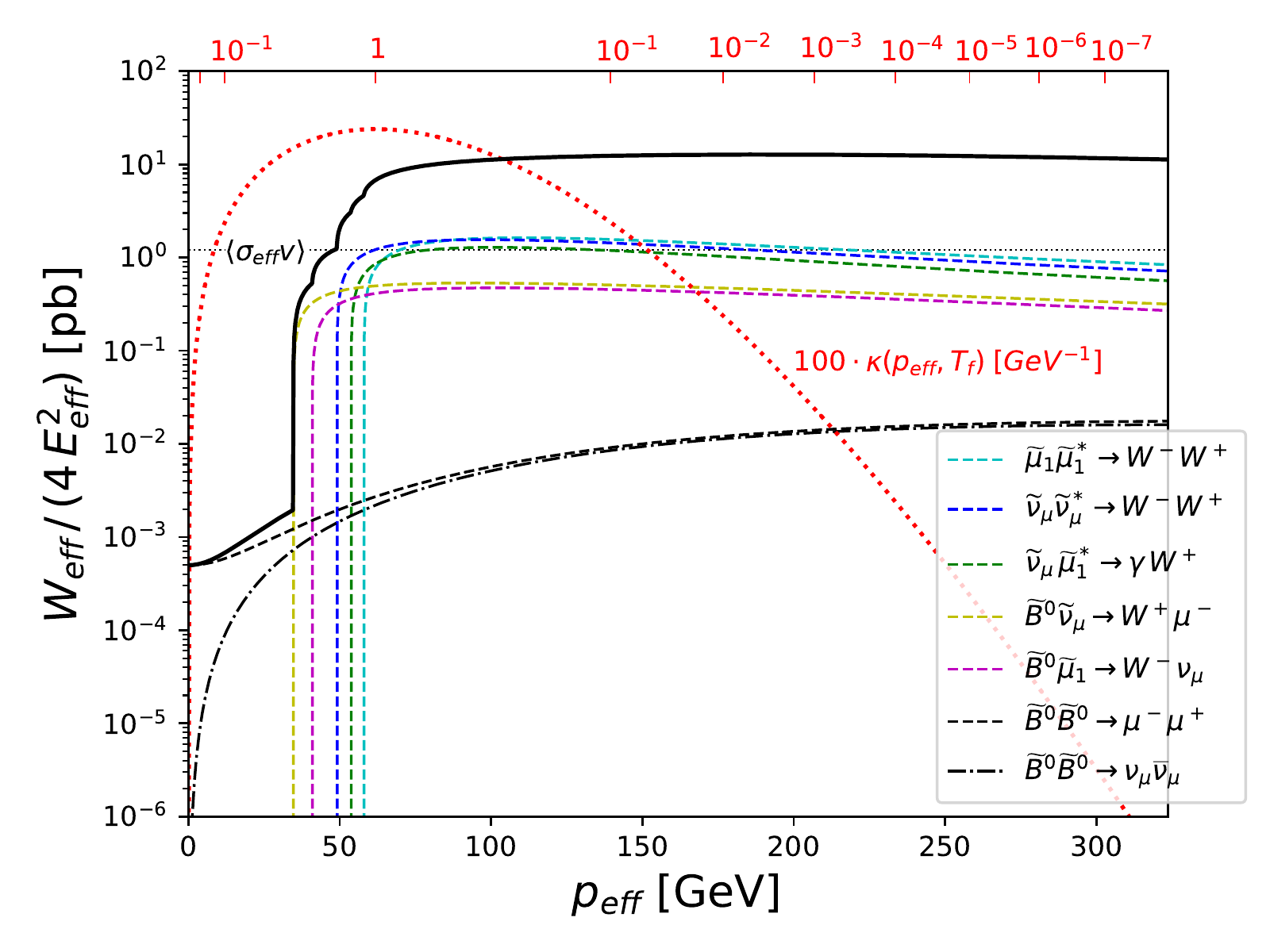}
\includegraphics[scale=0.54]{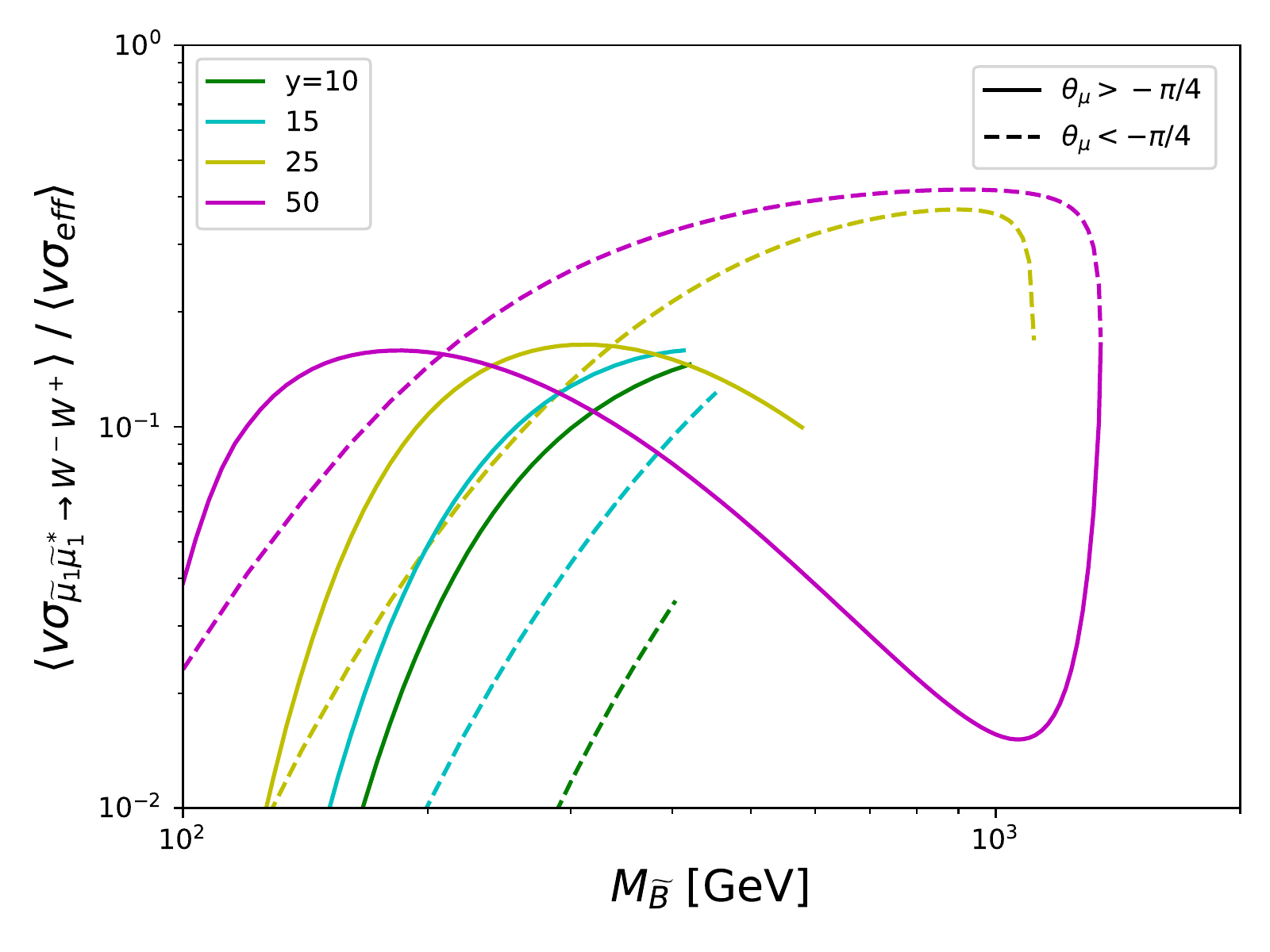}
\caption{{\sl Left panel:} The thick solid line is the effective annihilation cross section plotted versus the effective momentum for a sample DM model, see the text for details; also shown are individual contributions from the two tree-level final states in the Bino pair annihilation channel and from a single final state in each coannihilation channel, the one providing the largest contribution to the thermally averaged annihilation cross section. The dotted line is the weight function $\kappa$, computed at the freeze-out temperature and rescaled by a factor of 100, given in units of GeV$^{-1}$ still referring to the displayed vertical axis scale; the convolution of $W_{\rm{eff}}/ 4 E^2_{\rm{eff}}$ with $\kappa$ gives the thermally averaged effective annihilation cross section, shown in the plot as a horizontal thin dotted line. {\sl Right panel:} Relative weight of  $\tilde{\mu}_1 \tilde{\mu}_1^* \rightarrow W^-W^+$ in the thermally averaged effective annihilation cross section for a subset of models displayed in Fig.~\protect{\ref{fig:oh2iso1}}.
}
\label{fig:sigmaeff}
\end{figure}

In the left panel of Fig.~\ref{fig:sigmaeff} the thick solid line displays $W_{\rm{eff}}/4{E^2_{\rm{eff}}}$ versus $p_{\rm{eff}}$ for a sample model in our scans matching the $g_\mu-2$ excess and the relic density constraint. We consider a point with $M_{\tilde{B}} = 300 \,$GeV along the LHB at $y=15$, and relative mass splittings with the lightest smuon and the sneutrino being, respectively, 1.85\% and 1.33\%; the mixing angle is about  $\theta_{\tilde{\mu}} \simeq -$~\ang{1.6}. The effective rate picks up contributions from individual annihilation and coannihilation channels, with each contribution appearing at thresholds in $p_{\rm{eff}}$ corresponding the value of $\sqrt{s}$ equal to the sum of the masses of the initial state particles. Regarding individual terms, we display the two allowed tree-level final states for the Bino pair annihilations (with the $s$-wave contribution suppressed and the $p$-wave contribution taking over). For each coannihilation process, we display the final state providing the largest contribution to the thermally averaged annihilation cross section. The role of coannihilating particles is made explicit by the weight function $\kappa$, which is also displayed in the plot. $\kappa$ is plotted at the freeze out temperature, chosen here for illustrative purposes to be the temperature at which the abundance of the relic species is 50\% higher than the equilibrium value. At the top of the panel, the tick mark labelled `1' indicates the position of the momentum $p_{\rm{eff}}^{\rm{max}}$ corresponding to the maximum of $\kappa$, while the other tick marks indicate the momenta $p_{\rm{eff}}^{(n)}$ at which $\kappa(p_{\rm{eff}}^{(n)})/\kappa(p_{\rm{eff}}^{\rm{max}}) = 10^{-n}$. The tick marks provide a visual guide to the interval in $p_{\rm{eff}}$ which is relevant in the thermal averaging: the convolution of $W_{\rm{eff}}/ 4 E^2_{\rm{eff}}$ with $\kappa$ gives $\langle \sigma_{\rm{eff}}v \rangle$ thermally averaged at the freeze out temperature, shown in the figure as a horizontal thin dotted line and in fair agreement with what is expected based on the rule of thumb in Eq.~(\ref{eq:roft}). 

The sample model we have considered illustrates rather generic trends: There is no single coannihilating channel which is clearly driving the system of coupled Boltzmann equations, nor a single final state dominating the annihilation rate for a given pair of particles in the initial state; a slight change in any parameter of our model drives a ``coherent" shift for several terms. The other recurrent feature is the small contributions from Bino pair annihilation relative to all of the other terms, indicating the Bino relic density is extremely sensitive to the mass splittings between the Bino and the lightest scalars. In particular, from the point of view of the relic density calculation, a slight change in the Bino mass can be consistent with relatively large and compensatory changes of the parameters to which the coannihilation rates are most sensitive, specifically the mass splittings and $y$.

To investigate the peculiar relation between Bino mass and mass splitting with the coannihilating particle at moderate to large values of $y$ found in Figs.~\ref{fig:oh2iso1} and \ref{fig:oh2iso2}, we consider the process:
\be
\tilde{\mu}_i(p_1)+\tilde{\mu}_j^*(p_2)\rightarrow W^-(k_1)+W^+(k_2) \,.
\label{eq:wwfin}
\ee
For lightest smuons in the initial state ($i=j=1$), going to heavy masses and large $y$, this is one of channels with largest weights in the effective annihilation rate. The right panel of Fig.~\ref{fig:sigmaeff} shows the relative contribution of this single coannihilation process to the total effective annihilation rate for the models with the moderate to large values of $y$ already considered in Fig.~\ref{fig:oh2iso1}; it can reach a level of 40\% for $M_{\tilde{B}} \simeq 1 \,$TeV along the RHB. 

At any point in the parameter space considered here, the cross section for the process in Eq.~(\ref{eq:wwfin}) correctly scales as $\sigma_{ij} \propto 1/s$ for $s\rightarrow \infty$. On the other hand, the cross section is not $s$-wave suppressed and the limit that is most relevant for the effective thermally averaged annihilation rate is the one in which the three-momenta of the initial state particles $|\vec{p}_{1,2}|\rightarrow 0$. In this limit and for $M_{\tilde \mu_i},M_{\tilde \mu_j} \gg M_W$, one would generally expect that $\sigma_{ij} v$ scales as the inverse of the square of the masses in the initial state or, equivalently, that the amplitude squared scales a constant function of the smuon masses. This does not happen for a generic slicing of our parameter space: as we detail in the following, $y$ appears again, at different levels, as a crucial parameter. 
 
The expression of the modulus squared of the amplitude, summed over $W$ polarizations, for $|\vec{p}_{1,2}|\rightarrow 0$ takes the form
\be
 \sum_{\lambda \lambda^\prime} \left|{\mathcal M}_{ij} \right|^2 = g^2 \left\{
 \left[{\mathcal A}_{ij} \frac{(M_{\tilde \mu_i}+M_{\tilde \mu_j})^2}{4 M_W^2} 
- {\mathcal A}_{ij} + \frac{{\mathcal B}_{ij}}{2} \right]^2 +\frac{{\mathcal B_{ij}}^2}{2} \right\} \, ,
\label{eq:WWamp2}
\ee
where ${\mathcal A}_{ij}$ and ${\mathcal B}_{ij}$ are given in terms of contributions from the diagrams with $\tilde{\nu}_\mu$ in the $t$-channel, the four-point smuon-$W$ vertex, and $H^0$ in the $s$-channel (see Table~\ref{tab:coanns} and note that the diagrams with $\gamma$ and $Z^0$ in the $s$-channel do not contribute in the limit $|\vec{p}_{1,2}|\rightarrow 0$):
\bea
\begin{aligned}
& {\mathcal A_{ij}} = {\mathcal C}_{\tilde{\nu}_\mu ij}+ {\mathcal C}_{p ij} + {\mathcal C}_{H^0 ij}  \quad \quad {\mathcal B}_{ij} = {\mathcal C}_{p ij} + {\mathcal C}_{H^0 ij} \\
& {\mathcal C}_{\tilde{\nu}_\mu ij} = \frac{2 M_{\tilde \mu_i} M_{\tilde \mu_j}}{- M_{\tilde \mu_i} M_{\tilde \mu_j}-M_{\tilde{\nu}_\mu}^2+M_W^2} U_{iL} U_{jL} \quad \quad {\mathcal C}_{p ij} =U_{iL} U_{jL}
 \quad \quad {\mathcal C}_{H^0 ij} = \frac{2 M_W^2}{(M_{\tilde \mu_i} +M_{\tilde \mu_j})^2-M^2_{H^0}} y_{H^0 ij}
\end{aligned}
 \eea
with the Higgs coupling that, within the MSSM-like scheme introduced in Eq.~(\ref{eq:higgscoup}), is given by
\be
\label{yh0ij} y_{H^0 ij} \equiv \frac{1-\tan^2\theta_W}{2} U_{iL} U_{jL} + \tan^2\theta_W U_{iR} U_{jR} + y \,{\rm sgn}(\sin(2 \theta_{\tilde{\mu}})) (U_{iL} U_{jR}+U_{iR} U_{jL})\,,
\ee
and we have generically indicated with $U_{iL}$ and $U_{iR}$ the projection of the smuon $i$ on, respectively, the left- and right-handed fields. The expression in Eq.~(\ref{eq:WWamp2}) shows that the modulus squared of the amplitude would not increase with the smuon masses only if the inverse scaling with $M_W^2$ is cancelled out, \ie~if ${\mathcal A}_{ij}\propto M_W^2$ for $M_{\tilde \mu_i},M_{\tilde \mu_j} \gg M_W$ (there is no inverse scaling with $M_W^2$ in ${\mathcal B}_{ij}$). Since this is explicitly the case for the ${\mathcal C}_{H^0 ij}$ term, one only needs to examine the behaviour of the $\tilde{\nu}_\mu$ and four-point contributions. 

We do this check for the sample case of annihilation between the lightest smuons, \ie~when $i=j=1$; inserting the expression for the sneutrino mass Eq.~(\ref{eq:snumass}), one finds
\be
{\mathcal C}_{\tilde{\nu}_\mu 11}+ {\mathcal C}_{p 11} = \left\{1 - \frac{1}{1-(M^2_{W}+\Delta M^2_{W})/(2\,M_{\tilde \mu_1}^2)
+ |\tan(\theta_{\tilde{\mu}})| \,y\cdot M^2_{W}/M_{\tilde \mu_1}^2}\right\}  \cos^2(\theta_{\tilde{\mu}})\,.
\ee
Considering first a purely left-handed lightest smuon, $\theta_{\tilde{\mu}}=0$, one sees that ${\mathcal A}_{11}\propto M_W^2$ can be obtained only if the splitting $|\Delta M^2_{W}|$ in the sneutrino mass squared is not much larger than $M_W^2$, \ie~the sneutrino decoupling limit cannot be taken. For example, in the MSSM-like case with $\Delta M^2_{W} = M_W^2$, one finds ${\mathcal C}_{\tilde{\nu}_\mu 11}+ {\mathcal C}_{p 11} |_{\theta_{\tilde{\mu}}=0} = - M_W^2/M_{\tilde \mu_1}^2 + {\mathcal O} (M_W^4/M_{\tilde \mu_1}^4)$. Allowing for $\theta_{\tilde{\mu}}\ne 0$, on the LHB an analogous expansion can only be performed if $|\tan(\theta_{\tilde{\mu}}) \,y|$ does not become large, namely for moderate values of $y$. On the RHB, $|\tan(\theta_{\tilde{\mu}}) \,y|$ is large even for small $y$ and the cancellation between the leading terms of the sneutrino and 4-point diagrams does not take place. However the term is suppressed if the $\cos^2(\theta_{\tilde{\mu}})$ factor in the numerator is sufficiently small (the two diagrams are relevant only for left-handed interaction eigenstates). In summary, for moderate to large values of $y$ and sizable left-right mixing, the annihilation cross section for the lightest smuons (as well as other processes) can potentially grow as the masses of the coannihilating particles increase. This peculiar effect can be particularly important for models fulfilling $g_\mu-2$ and the relic density in Fig.~\ref{fig:oh2iso1}, but is also relevant for the models shown in Fig.~\ref{fig:oh2iso2}.

While this statement holds regardless of what is assumed for the coupling $y_{H^0 ij}$, within our MSSM-like scheme a large Higgs contribution to the amplitude is also present for sizable left-right mixing and large $y$, simply because both enter linearly in $y_{H^0 ij}$. More specifically, we can go back to Eq.~(\ref{yh0ij}) and consider the case where the last term dominates the $s$-channel Higgs contribution to the amplitude for $\tilde{\mu}_i \tilde{\mu}_j^* \rightarrow W^-W^+$, \ie
\begin{eqnarray} \label{eqn:yH0ij}
y_{H^0 ij} \approx y~\text{sgn}(\sin(2\theta_{\tilde{\mu}}))S_{xij}(\theta_{\tilde{\mu}}),\quad S_{x}(\theta_{\tilde{\mu}}) \equiv \left(\begin{matrix}
-\sin(2\theta_{\tilde{\mu}})&\cos(2\theta_{\tilde{\mu}})\\
\cos(2\theta_{\tilde{\mu}})&\sin(2\theta_{\tilde{\mu}})
\end{matrix}\right).
\end{eqnarray}
Again focusing on the annihilation between the lightest smuons, $i=j=1$, we have $y_{H^0 11} \approx -y\vert\sin(2\theta_{\tilde{\mu}})\vert$. If we then consider $M_{\tilde \mu_1}^2 \gg M^2_{H^0}$, we see there is a contribution to the amplitude squared in Eq.~(\ref{eq:WWamp2}) $\propto y^2 \sin^2 (2\theta_{\tilde{\mu}})$ arising from the corresponding ${\mathcal C}_{H^0 11}$ term in ${\mathcal A}_{11}$. Regarding the associated cross section for smuon annihilation in the limit $|\vec{p}_{1,2}|\rightarrow 0$, there is a suppression of this term as $M_{\tilde \mu_1}$ increases rather than the peculiar growth seen when considering the spoiled cancellation between gauge interactions described above. However, for fixed $M_{\tilde \mu_1}$, this term from the $s$-channel Higgs contribution to the amplitude can raise the smuon annihilation cross section as the mixing angle is maximized and $y$ is taken to be arbitrarily large. This scaling explains much of the relationship between $y$ and the mass splitting shown for models fulfilling $g_\mu-2$ and the relic density in Fig.~\ref{fig:oh2iso2}. In addition, as one can see in the right panel of Fig.~\ref{fig:sigmaeff}, on the RHB the Higgs diagram drives a further enhancement to the smuon annihilation rate, while on the LHB there is a partial cancellation between the gauge and Higgs contributions. Such cancellation is accidental for the particular process of the lightest smuons annihilating to $W$-bosons and cross sections in this limit remain potentially problematic for other coannihilation channels, \eg~for the same initial state and two $Z^0$ bosons in the final state.

As a rule of thumb, we could exclude models with too large couplings/cross sections by implementing limits imposed by requiring the unitarity of partial wave cross sections for individual contributions to the effective annihilation cross section~\cite{Griest:1989wd},
\begin{eqnarray}
\label{inequality}(\sigma v)_J \leq \frac{4\pi (2J+1)}{m_i^2 v}\, ,
\end{eqnarray}
where $J$ is the angular momentum and $m_i$ is some common mass of initial state particles. In practice, all models displayed in Figs.~\ref{fig:oh2iso1} and \ref{fig:oh2iso2} do not violate such bounds. However, models with large $y$ and sizable mixing angles are severely constrained when considering the unitarity of the general form of the scattering matrix. Even more stringent constraints arise from considering the vacuum structure in our theory and the (meta)stability of the EW vacuum. We will discuss these issues in the next Section.

\section{Perturbative unitarity and vacuum stability} \label{sec:pert}
\noindent
The analysis at the end of the previous section regarding the squared amplitude of the process $\tilde{\mu}_i \tilde{\mu}_j^* \rightarrow W^-W^+$ suggests that $y$ should not exceed a certain value for a fixed $\theta_{\tilde{\mu}}$. This condition on $y$ follows from the requirement that the squared amplitude be sufficiently small for annihilation and coannihilation processes which deplete the relic density, such that we are well within the regime for which perturbativity still holds. Thus, requiring perturbative unitarity can impose theoretical limits on the allowed couplings and masses of the species in our setup. In particular, if we assume the MSSM-like benchmark for the Higgs trilinear couplings with the sleptons introduced in Eq.~(\ref{eq:higgscoup}), theoretical limits on $y$ correspond to constraints on the off-diagonal coupling between the Higgs and the smuons. Such a criterion based on the simple principle of perturbative unitarity has been used, for instance, in Ref.~\cite{LEE1991282} to obtain an upper bound on the Higgs mass, long before its discovery. 

The key concept behind the determination of constraints from perturbative unitarity is the condition that one imposes on the $J=0$ partial wave amplitude. We begin by writing down the condition on the transition matrix elements $T_{fi}$ from the unitarity of the S-matrix, \ie
\begin{eqnarray}
\label{opticaltheorem}\Im\{T_{fi}\} = \sum_k T_{kf}^* T_{ki}.
\end{eqnarray}
We only consider the block of the S-matrix that corresponds to two-particle initial ($i$) and final ($f$) states, \ie~processes of type $\phi_1 \phi_2 \rightarrow \phi_3 \phi_4$. The sum in Eq.~(\ref{opticaltheorem}) runs over all possible intermediate states and quartic interaction terms $k$. For simplicity, we only consider two-scalar initial and final states, as well as scalar mediators in the sum on $k$; this underestimates the right-hand side of Eq.~(\ref{opticaltheorem}), which leads to conservative bounds. In general, the entries in the S-matrix depend on the center of mass energy $\sqrt{s}$ and the scattering angle $\theta$, which can be traded with the Mandelstam variable $t$; the tree level amplitude for a 2-2 process $\phi_1 \phi_2 \rightarrow \phi_3 \phi_4$ can be heuristically written as 
\begin{eqnarray}
T_{fi} = c_4 + \frac{c_s}{s-m_{P,s}^2} + \frac{c_t}{t-m_{P,t}^2} + \frac{c_u}{u-m_{P,u}^2},
\end{eqnarray}
where $c_4$, $c_s$, $c_t$, and $c_u$ are quantities, with the appropriate mass dimension, that are built from the couplings in the theory. It is then convenient to sift out the angular dependence of the scattering amplitudes by implementing a partial wave decomposition. By projecting the transition amplitudes on a complete set of Legendre polynomials $P_J(\cos\theta)$, it can be shown that
\begin{eqnarray}
\label{opticaltheoremJ}2\Im\{a_{fi,J}\} \leq \sum_k a_{kf,J}^* a_{ki,J},
\end{eqnarray}
for all $J$ \cite{goodsell2018unitarity}. Here the partial wave matrix element $a_{fi,J}$ is given by
\begin{eqnarray}
\label{partialwaveJ}a_{fi,J}(s) \equiv \frac{1}{32\pi} \sqrt{\frac{4 p_1 p_3}{2^{\delta_{12}} 2^{\delta_{34}} s}} \int_{-1}^1 d(\cos\theta)~T_{fi}(s,\cos\theta)~P_J(\cos\theta),
\end{eqnarray}
where $p_1 (p_3)$ is the magnitude of the 3-momenta of the initial (final) states in the barycentric frame, and $\delta_{12} (\delta_{34})$ is zero if particles 1 and 2 (3 and 4) are nonidentical, and 1 otherwise. 
Further restricting our attention to the $J=0$ partial wave, and diagonalizing $a_0$, Eq.~(\ref{opticaltheoremJ}) implies that the eigenvalues $a_0^{(i)}(s)$ must satisfy
\begin{eqnarray}
\label{equationcircle}\Im\{a_0^{(i)}(s)\} \leq \left\vert a_0^{(i)} \right\vert^2 \Rightarrow \left[\Re\{a_0^{(i)}(s)\}\right]^2+\left(\Im\{a_0^{(i)}(s)\}-\frac{1}{2}\right)^2 \leq \frac{1}{4}.
\end{eqnarray}
We emphasize that Eq.~(\ref{equationcircle}) holds at all orders in perturbation theory, since there is no assumption that the amplitudes are truncated at tree level. In the case where Eq.~(\ref{equationcircle}) is an equality---an assumption used in Refs.~\cite{betre2014perturbative} and~\cite{Schuessler:2007av}---Eq.~(\ref{equationcircle}) determines the so-called unitarity circle in the complex $\hat{a}_0^{(i)}(s)$ plane; any transition amplitude that satisfies unitarity must lie on this circle. However, at tree level---the order at which all of the cross sections, \eg~for the relic density, are calculated---the transition matrix is real and symmetric, the eigenvalues are always real, and thus the partial wave, tree level amplitude will always lie outside the unitarity circle. In principle, one will approach the unitarity circle if one includes corrections from all orders in perturbation theory, including loop contributions, to the amplitude \cite{betre2014perturbative}. An estimate of the amount of loop corrections to the tree level amplitude, in order to satisfy unitarity, can be obtained by taking the closest distance $d$ between the unitarity circle and the tree level amplitude that lies on the real axis. 

The criterion that one can adopt to ensure perturbative unitarity is to set 
\begin{eqnarray}
a \equiv \frac{d}{\text{max}\{\vert \Re\{a_0^{(i)}(s)\}\vert\}} = \frac{\sqrt{1+4\lambda_{max}^2} - 1}{2\lambda_{max}}
\end{eqnarray}
to be less than some value which ensures the unitarity of the scattering matrix. Here, $i$ runs over all the eigenvalues of the partial wave S-matrix. Note that the maximum eigenvalue $\lambda_{max}$ is taken over all partial wave S-matrix eigenvalues and over all physically allowed $s$. It is worth mentioning that we are implicitly pointing out that the strongest limits from perturbative unitarity can occur at finite energy in theories with large trilinear couplings, in contrast with some previous works, \eg~\cite{Hartling:2014zca,Khan:2016sxm}, that study limits on quartic couplings in the $s \to \infty$ limit. Ref.~\cite{Schuessler:2007av} adopts $\vert \Re\{\lambda_{max}\} \vert \leq 1/2$ for perturbative unitarity, and $\vert \Re\{\lambda_{max}\} \vert \leq 1/6$ to ensure the smallness of the Born amplitude; each criterion corresponds, respectively, to at most 41\% and 16\% corrections from higher orders to ensure unitarity.

Before turning to the constraints that requiring perturbative unitarity can place on our model, we first describe the full scalar potential in detail. In addition to the mass term for the sleptons and the trilinear Higgs-slepton couplings specified in Eq.~(\ref{eq:higgscoup}), we also must include quartic terms in order for the potential to be bounded from below. From the perspective of effective field theory, all quartic interactions allowed by the symmetries of the Lagrangian should be included in the scalar potential. As a benchmark, we consider quartic interactions and couplings arising from the D-term of the scalar potential in the MSSM. As discussed in Sec.~\ref{sec:model}, such terms do not significantly impact the observables in our simplified model, but can be important for the constraints on our model arising from perturbative unitarity and EW vacuum stability. The corresponding terms in the full tree-level scalar potential are given by
\begin{eqnarray}
V_2 &=& m_{LL}^2 \tilde{l}_L^\dag \tilde{l}_L + m_{RR}^2 \tilde{\mu}_R^\dag \tilde{\mu}_R + \mu^2 H^\dag H\\
V_{mix} &=& k_s \left(H^\dag \tilde{l}_L \tilde{\mu}_R^\dag+ \tilde{l}_L^\dag H \tilde{\mu}_R\right)\\
V_D^{(1)} &=& \frac{{g^\prime}^2}{2}\left\vert Y_H H^\dag H + Y_L \tilde{l}_L^\dag \tilde{l}_L + Y_R \tilde{\mu}_R^\dag \tilde{\mu}_R\right\vert^2\\
V_D^{(2)} &=& \frac{g^2}{4}\left\{\text{tr}\left(M^2\right) - \frac{1}{2}\left[\text{tr}(M)\right]^2\right\} \, ,
\end{eqnarray}
where $\mu^2$ is the Higgs mass parameter\footnote{To recover the
EW vacuum of the SM, we see that the potential has a minimum along the field direction of the physical Higgs boson for $\mu^2 < 0$. Note that the associated tadpole condition fixes the Higgs mass to the $Z^0$-boson mass at tree-level in the Lagrangian, as for the MSSM in the limit where the vacuum expectation value (VEV) for one of the Higgs doublets vanishes. For simplicity, when calculating observables predicted by our model we assume that some additional mass contributions (\eg~loops of additional scalars) raise the Higgs mass to what is observed by LHC, $M_{H^0} \simeq \unit[125]{GeV}$. On the other hand, as we only calculate the constraints from perturbative unitarity and vacuum stability using the tree-level potential, we assume $M_{H^0} = M_{Z^0}$ when evaluating constraints on our model from theoretical consistency.}, $k_s \equiv \sqrt{2} g M_W y$ and $M \equiv H H^\dag + \tilde{l}_L \tilde{l}_L^\dag $. The total tree-level potential can then be written as
\begin{eqnarray} \label{eq:scalpot}
V_{tot} =  V_2 + V_{mix} + V_D^{(1)} + V_D^{(2)} \, .
\end{eqnarray}
Note that the analysis of perturbative unitarity in the S-matrix proceeds in the physical basis of mass eigenstates and, thus, the interactions involving the smuons in the scalar potential above should be considered in terms of the eigenstates arising from the diagonalization of the smuon mass matrix described in Sec.~\ref{sec:model}. In order to account for processes in the S-matrix involving gauge bosons, such as $\tilde{\mu}_i \tilde{\mu}_j^* \rightarrow W^-W^+$, we work in the Feynman $R_{\xi=1}$-gauge. The Goldstone bosons associated with the SM-like Higgs doublet then represent the longitudinal polarizations of the SM gauge bosons and, by the Goldstone boson equivalence theorem, each has a mass equivalent to the corresponding gauge boson (for related discussion see~\cite{Schuessler:2007av,goodsell2018unitarity}). For the analysis of vacuum stability below, it is more convenient to work in the chiral basis of the smuons and the unitary gauge.

At this point we can discuss the practical aspects of the perturbative unitarity analysis for the parameter space of our model which can satisfy both $g_\mu - 2$ and the relic density. We scan along curves of constant $y$ or constant $\theta_{\tilde{\mu}}$ shown in Figs.~\ref{fig:oh2iso1} and~\ref{fig:oh2iso2}, respectively. For the curves with fixed $y$, perturbative unitarity sets a limit on the smuon mixing angle or, through the dependence of $\theta_{\tilde{\mu}}$ on $M_{\tilde{B}}$ necessary to satisfy $g_\mu - 2$ shown in the left panel of Fig.~\ref{fig:thetamu}, the masses of the coannihilating particles. For the curves with fixed $\theta_{\tilde{\mu}}$, perturbative unitarity sets a limit on $y$ or, equivalently, on the masses of the coannihilating particles implied by the dependence of $M_{\tilde{B}}$ on $y$ in the right panel of Fig.~\ref{fig:thetamu}.

For each parameter point, we construct the $J = 0$ partial wave projection of the S-matrix from tree level amplitudes for all 2-particle initial and final states possible in the scalar potential described above. In principle, the maximum eigenvalue of the partial wave S-matrix is obtained by scanning over all physically allowed $\sqrt{s}$. Through this process, we extract the maximum eigenvalue $\lambda_{max}$ as well as the center of mass energy at which this maximum eigenvalue occurs, which we refer to as the \textit{best energy}. However, there are subtle points that must be addressed when performing this scan in $\sqrt{s}$, particularly in handling the poles associated with propagators that go on shell. In previous studies of pertrurbative unitarity~\cite{goodsell2018unitarity,goodsell2019improved}, a pole cutting procedure is implemented in order to avoid artificial enhancements to the S-matrix elements arising from physical poles. However, such a procedure may unnecessarily prune out some portions of the scan in $\sqrt{s}$, which could result in an underestimate of the matrix elements containing the poles and weakened unitarity limits. The specific implementation of the pole cutting procedure could also not be sufficient to completely eliminate any enhancements to S-matrix elements associated with the physical poles, resulting in overly stringent unitarity limits. 

Rather than implementing a similar pole-cutting procedure in our analysis, we regulate the singular behavior of poles by introducing an artificial width $\Gamma_i = b_i m_P$ to each propagator with mass $m_P$; here $i$ refers to the width for $s$-, $t$-, or $u$-channel propagators. We then restrict our scans in $\sqrt{s}$ within the so-called \textit{safe intervals} for the S-matrix of a given parameter point. Due to the characteristic scaling of all S-matrix elements $\propto 1/s$, the best energy is typically located near the kinematic threshold of a certain process. We therefore define safe intervals to be bounded from below in $\sqrt{s}$ by the kinematic thresholds for all possible 2-particle states in our model and bounded from above by some constant multiple of each kinematic threshold, $1+\eta$. We then ensure that each safe interval does not overlap with intervals in $\sqrt{s}$ that are centered about the physical poles. We shall refer to these as \textit{pole intervals}; for center of mass energy associated with each pole $\sqrt{s_*}$, the pole interval is defined as $[\sqrt{s_*}(1-\epsilon),\sqrt{s_*}(1+\epsilon)]$. For any safe interval not associated with the highest kinematic threshold which does overlap with a pole interval, we remove the safe interval from our scan of the S-matrix. We keep the safe interval associated with the highest kinematic threshold in all scans of $\sqrt{s}$ since that interval typically contains the largest eigenvalue not enhanced by a physical pole.\footnote{We leave a detailed comparison of the different techniques used to analyze perturbative unitarity to future work, which will also include an investigation of semi-analytic approximations for the bounce action discussed below in the context of EW vacuum stability.}

\begin{figure}[t!]
\centering
\includegraphics[scale=0.54]{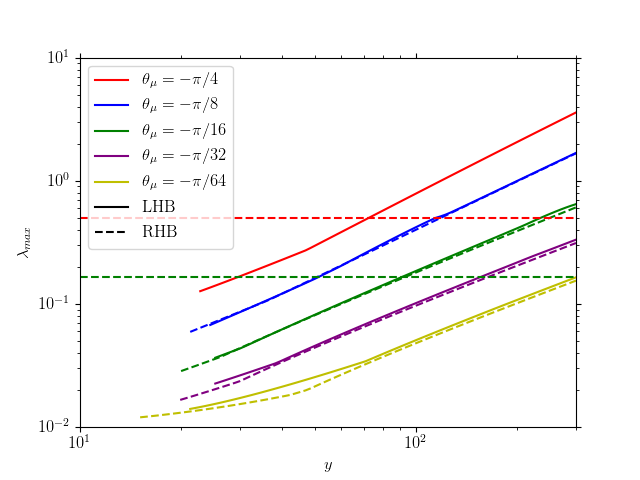}\quad \includegraphics[scale=0.54]{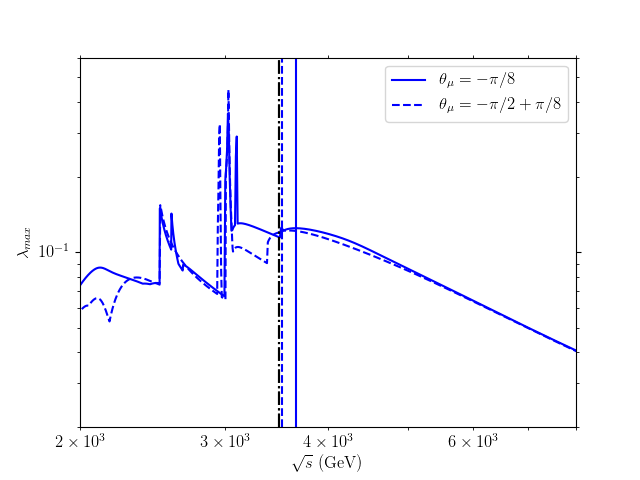}
\caption{\label{fig:pertunitscanisoth} {\sl Left panel:} Plot of the maximum S-matrix eigenvalue as a function of $y$, along curves of constant $\theta_{\tilde{\mu}}$ where both $g_\mu -2$ and the relic density are satisfied. The LHB (solid curves) corresponds to $\theta_{\tilde{\mu}}$, while the RHB (dashed curves) corresponds to $-\pi/2-\theta_{\tilde{\mu}}$. The horizontal green and red dashed lines correspond to upper limits of 1/6 and 1/2, respectively. {\sl Right panel:} Scans in $\sqrt{s}$ for a benchmark value of $y \simeq 41$, for fixed $\theta_{\tilde{\mu}}$ values $-\pi/8$ and $-\pi/2+\pi/8$ in the LHB and RHB, respectively. The black dot-dashed line corresponds to the $\sqrt{s}$ of the maximum possible threshold, while the blue vertical lines correspond to the values of $\sqrt{s}$ which gives the maximum eigenvalue across all safe intervals.}
\end{figure}

The algorithm outlined above can be efficiently implemented by first specifying our model in \verb|SARAH| \cite{staub2014sarah}, and then generating the associated \verb|SPheno| \cite{porod2003spheno,porod2012spheno} code for scans of the S-matrix. We have modified the \verb|SPheno| code to accommodate for the widths in the $s/t/u$-channel propagators and we have not implemented any of the available pole cutting procedures. We fix the coefficients of the widths to be $b_s = b_t = b_u = 0.5$ and define the safe and pole intervals with the parameters $\eta = 0.25$, $\epsilon = 0.1$. In the left panel of Fig. \ref{fig:pertunitscanisoth} we show the maximum eigenvalues of the S-matrix as a function of $y$ along the curves of constant $\theta_{\tilde{\mu}}$ from Fig.~\ref{fig:oh2iso2}, for which $g_\mu - 2$ and the relic density limits are satisfied. The solid curves in Fig. \ref{fig:pertunitscanisoth} correspond to smuon mixing angles on the LHB, $\theta_{\tilde{\mu}} = -\pi/n$, $n = 4,~8,~16,~32,$ and $64$; the dashed curves refer to smuon mixing angles on the RHB, $\theta_{\tilde{\mu}} = -\pi/2 + \pi/n$, $n = 8,~16,~32,$ and $64$. For each iso-$\theta_{\tilde{\mu}}$ curve, $\lambda_{max}$ increases with $y$; meanwhile, for fixed $y$, $\lambda_{max}$ increases as we move towards maximal mixing, \ie~$\theta_{\tilde{\mu}} \rightarrow -\pi/4$ from either the LHB or RHB.

Both of these trends can be explained by the dominant contributions to the S-matrix, which can be traced back to processes that involve the ``dangerous" trilinear terms in the scalar potential $\propto y$. The dominant S-matrix element can involve one or two of these vertices, which is enhanced by either increasing $y$ for fixed $\theta_{\tilde{\mu}}$ or $\theta_{\tilde{\mu}}$ approaching maximal mixing for fixed $y$. This functional dependence of the S-matrix elements is consistent with the scaling of the corresponding annihilation and coannihilation rates discussed at the end of Sec.~\ref{sec:rel}. As depicted in the right panel of Fig.~\ref{fig:pertunitscanisoth} for a representative case, the best energy located within a safe interval is typically above the maximum possible threshold $2 M_{\tilde \mu_2}$ for a given spectrum. This suggests that the most important contribution ultimately comes from S-matrix elements with couplings like $y_{H^0 22} \approx y\vert\sin(2\theta_{\tilde{\mu}})\vert$, as defined by Eq.~(\ref{eqn:yH0ij}). Clearly, this coupling increases with $y$ for fixed $\theta_{\tilde{\mu}}$ and reaches a maximum at fixed $y$ when $\theta_{\tilde{\mu}} \rightarrow -\pi/4$. Also, since $\vert y_{H^0 22}(-\pi/4+\beta)\vert = \vert y_{H^0 22}(-\pi/4-\beta)\vert$, the dominant S-matrix element on the LHB is the same as that on the RHB. This match can be seen in the left panel of Fig. \ref{fig:pertunitscanisoth}, where the curves for $\theta_{\tilde{\mu}}$ and $-\pi/2 - \theta_{\tilde{\mu}}$ are nearly identical.

Additional constraints can be placed on the trilinear coupling in our model by considering the (meta)stability of the EW vacuum. The EW vacuum is said to be \textit{absolutely stable} if it corresponds to the global minimum of the potential. Otherwise, the EW vacuum is said to be \textit{metastable} and tunneling to the true vacuum of the theory will occur over some time scale which should be sufficiently long relative to the age of the Universe. This kind of analysis has been implemented in, \eg~\cite{hollik2019impact,duan2019vacuum}; in particular, Ref.~\cite{duan2019vacuum} shows that the EW vacuum can be sufficiently long-lived in the MSSM with maximally mixed staus as heavy as $\sim \unit[1]{TeV}$.

If we return to the scalar potential in Eq.~(\ref{eq:scalpot}), we can work in the unitary gauge and perform SU(2) rotations on the Higgs and left-handed slepton doublet. Any scalar field $\phi$ can be written in terms of its real and imaginary parts, such that
\begin{eqnarray}
\phi = \frac{1}{\sqrt{2}}\phi_R + \frac{i}{\sqrt{2}}\phi_I;
\end{eqnarray}
the normalization factor $1/\sqrt{2}$ ensures that we have canonical kinetic terms for $\phi_R$ and $\phi_I$. We assume CP-conservation in the scalar potential and a phase rotation on a field does not change the total potential, and hence we can simply work with the real parts of the different scalar fields. Letting $h$ to be the real component of the neutral Higgs, $X$ and $Y$ to be the real parts of $\tilde{\nu}$ and $\tilde{\mu}_L$, respectively, and $Z$ to be the real part of $\tilde{\mu}_R$, we have
\begin{eqnarray}
 V_2 &=& \frac{m_{LL}^2}{2}\left(X^2+Y^2\right)+\frac{m_{RR}^2}{2}Z^2 + \frac{\mu^2}{2}h^2 \\
V_{mix} &=& \frac{k_s}{\sqrt{2}}hYZ \\
V_D^{(1)} &=& \frac{{g^\prime}^2}{32}\left[h^2-\left(X^2+Y^2\right)+2Y_R Z^2\right]^2  \\
 V_D^{(2)} &=& \frac{g^2}{32}\left[h^4-2h^2\left(X^2-Y^2\right)+\left(X^2+Y^2\right)^2\right] \, .
\end{eqnarray}
With the relevant form of the scalar potential in hand, we now consider the vacuum structure. The EW vacuum corresponds to the minimum of the potential which develops when only the real part of the Higgs acquires a VEV, $v_{EW} = \unit[246]{GeV}$. When the real parts of the scalar fields other than the Higgs also acquire VEVs,  the potential can develop additional extrema due to the trilinear term. In particular, for the large trilinear couplings necessary to satisfy muon $g-2$ and the relic density, the global minimum of the potential corresponds to vacuum configurations where the Higgs and smuons acquire VEVs $\gg v_{EW}$. In addition to the global minimum, the trilinear term coupling the smuons to the Higgs also gives rise to a saddle point (SP) in the scalar potential. As we shall see, consideration of this saddle point is important when calculating the tunneling rate between vacua. Note that the sneutrino VEV vanishes for any vacuum configuration associated with the above potential and, thus, we only consider field trajectories involving the Higgs and smuons (\ie~$X=0$) in the analysis that follows.

The probability of tunneling from the EW vacuum to the global minimum of the tree-level potential at zero
temperature is given by \cite{hollik2019impact}
\begin{eqnarray}
P = \exp\left(-\mathcal{M}^4 \tilde{V}_{\text{light-cone}}e^{-B}\right)
\end{eqnarray}
where $\mathcal{M}$ is a characteristic scale of the theory and the spacetime volume of the past light-cone can be written in terms of the current value for the Hubble parameter $V_{\text{light-cone}} \sim 0.15/H_0^4$. While a precise determination of $\mathcal{M}$ is beyond the scope of this work, Ref.~\cite{hollik2019impact} demonstrates that the the EW vacuum can be considered metastable over timescales longer than the age of the universe for $B \gtrsim 440$ after considering a range of $\mathcal{M}$ within several orders of magnitude of the EW scale, $\sim \unit[1]{TeV}$. The four dimensional Euclidean bounce action is
\begin{eqnarray} \label{eq:bounceA}
B = \int_0^\infty d\rho \left[\mathcal{T} + \mathcal{V}\right], \mathcal{T} = \frac{\pi^2}{2}\rho^3\left[\frac{1}{2}\sum_{\phi=h,Y,Z}\left(\frac{d\phi}{d\rho}\right)^2\right], \mathcal{V} = \frac{\pi^2}{2}\rho^3 V_{tot}(h,Y,Z)	
\end{eqnarray}
for bounce solutions which are functions of the Euclidean radius, $\rho^2 = \sum_i x_i^2 - t^2$, along the field trajectories between the true vacuum (TV) and false vacuum (FV). For field configurations which extremize the bounce action, the Euclidean equation of motion and boundary conditions for each field are given by
\begin{eqnarray}
\label{eqmotion}\frac{d^2 \phi}{d\rho^2} + \frac{3}{\rho}\frac{d\phi}{d\rho} = \frac{\partial V_{tot}}{\partial \phi}, \phi(0) = \phi_{TV}, \phi(\infty) = \phi_{FV}, \frac{d\phi}{d\rho}(0) = \frac{d\phi}{d\rho}(\infty) = 0.
\end{eqnarray}
In our specific case with three relevant fields, the equations of motion are analogous to a particle moving through
a 3D potential. However the potential is inverted relative to the total potential (\ie~$-V_{tot}$ ) and the particle is also subject to a path-dependent drag term, which becomes singular as $\rho \rightarrow 0$. In practice, the calculation of the field configuration which minimizes the bounce action is often treated as boundary value problem where an initial ansatz for the field trajectory begins near the inverted global minimum of the potential and the bounce solution is iterated for different initial conditions until finding a field trajectory which ends on top of the inverted false minimum.

We calculate the bounce action for tunneling out of the EW vacuum using \verb|FindBounce| \cite{Guada:2020xnz}. The package implements a semi-analytical calculation of the bounce solution by first discretizing the potential into an interconnected series of finite, linear segments along the initial ansatz for the field trajectory. The associated polygonal bounce solution is then constructed by solving for the field trajectory along each segment, requiring the piecewise function to be continuous and differentiable at each segmentation point. Subsequently, the bounce solution can be perturbatively improved by expanding the potential to higher order at each segmentation point and iteratively building upon the polygonal bounce solution. \verb|FindBounce| is particularly well suited for the potentials we consider since the large trilinear terms necessary to satisfy $g_\mu-2$ and the relic density are associated with bounce solutions which deviate significantly from straight lines in field space connecting the true and false minima.

\begin{figure}[t!]
\begin{tabular}{cc}
\includegraphics[scale=0.445]{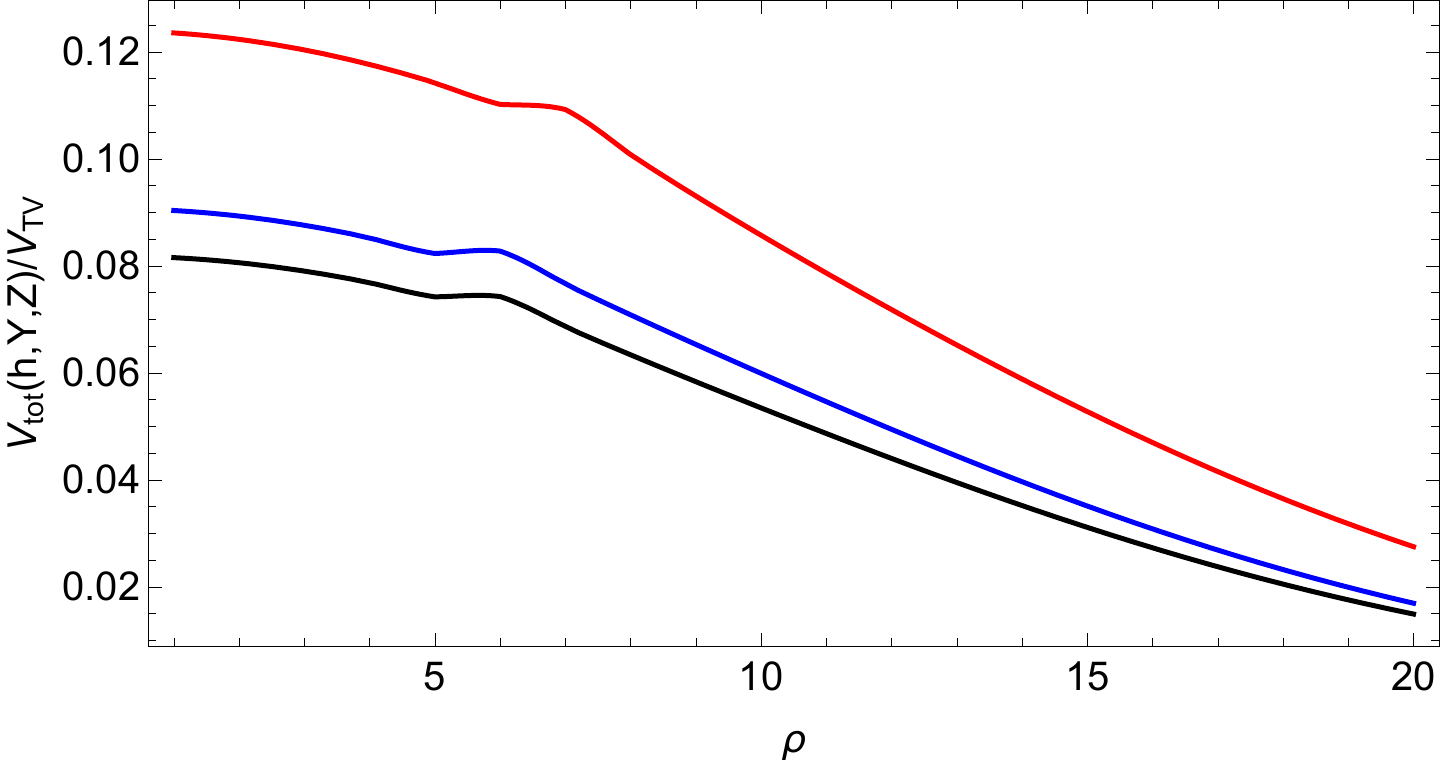} & \quad\multirow{2}{*}[1.25in]{\includegraphics[scale=0.5]{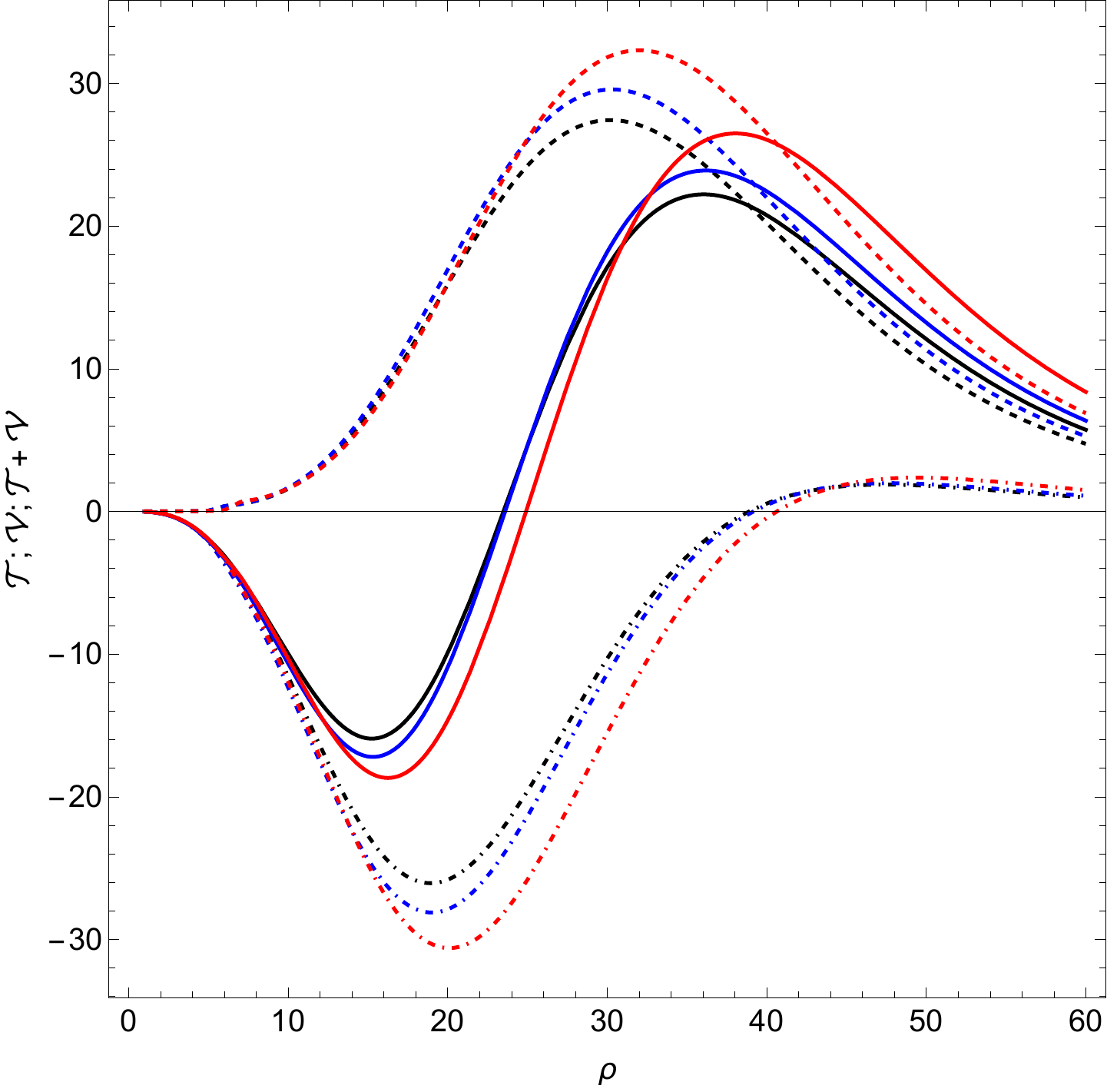}} \\
\includegraphics[scale=0.445]{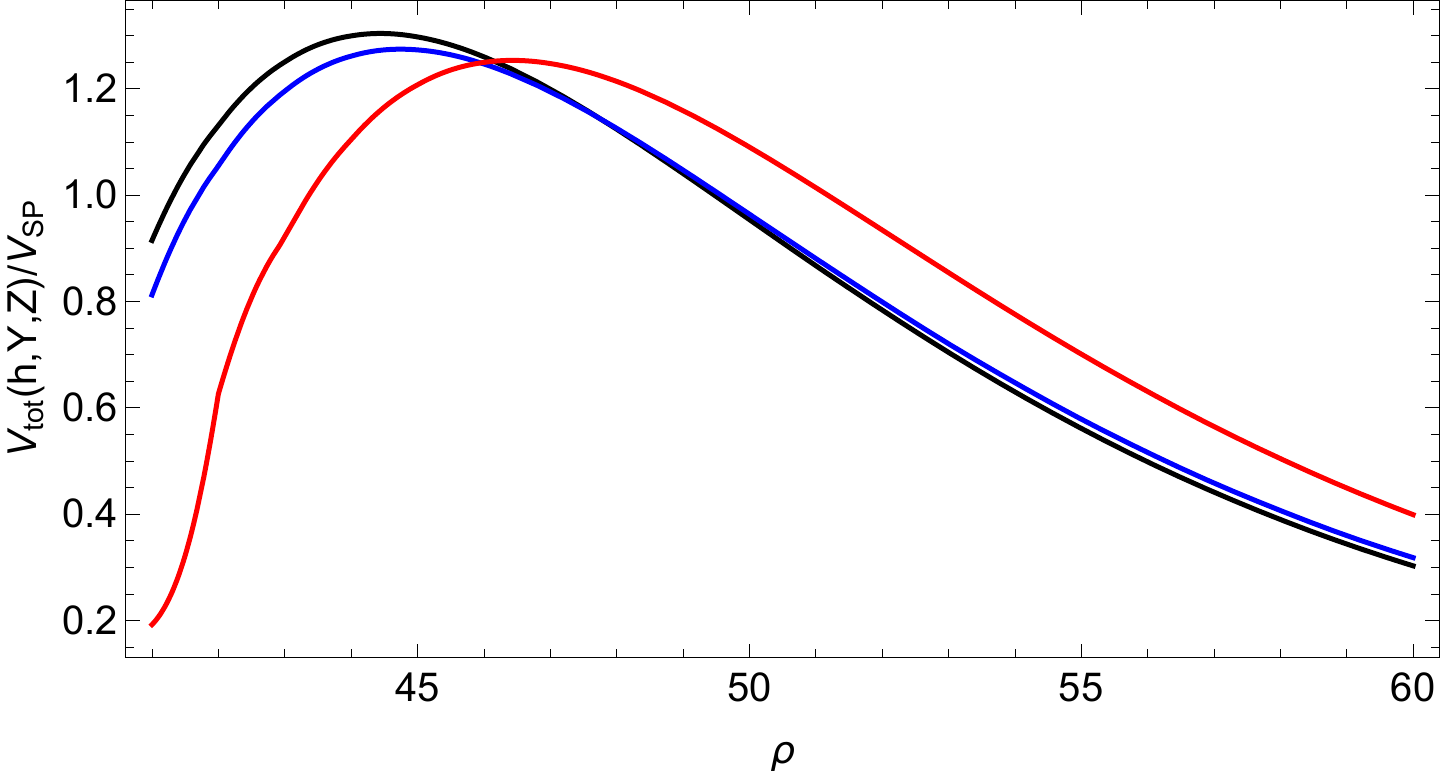} &                   
\end{tabular}
\caption{{\sl Left panels:} Plots of the potential as a function of Euclidean radius $\rho$ along the field trajectories corresponding to the bounce solutions for points shown in Fig.~\ref{fig:oh2iso2} with $\theta_{\tilde{\mu}} = -\pi/2+\pi/8$ and $y \simeq 23$ (black lines, $B \simeq 432$); $\theta_{\tilde{\mu}} = -\pi/2+\pi/16$ and $y \simeq 23$ (blue, $B \simeq 469$); $\theta_{\tilde{\mu}} = -\pi/2+\pi/8$ and $y \simeq 21$ (red, $B \simeq 530$). We show the potential at the beginning of the trajectory (normalized to its value at the TV) for each bounce solution in the top panel, while we show the potential along the field trajectory near the SP (normalized to its value at the SP) in the bottom panel. {\sl Right panel:} Plot of the separate contributions from the kinetic energy (dashed lines) and potential (dot-dashed) to the integrand for the bounce action, along with the sum (solid). The line colors correspond to the same models as in the left panel. Note that we have rescaled the kinetic term, potential and the Euclidean radius in Eq.~(\ref{eq:bounceA}) to be dimensionless and the bounce solutions all converge to the EW vacuum at $\rho \simeq 270$.}
\label{fig:bsol} 
\end{figure}
In the left panels of Fig.~\ref{fig:bsol}, we plot the value of the scalar potential along the field trajectories corresponding to the bounce solutions for several points from Fig.~\ref{fig:oh2iso2} which satisfy both $g_\mu - 2$ and the relic density. The black curves correspond to the bounce solution for the model with $\theta_{\tilde{\mu}} = -\pi/2+\pi/8$ and $y \simeq 23$, for which the bounce action is $B \simeq 432$. The model for the bounce solution shown by the blue curves holds $y$ constant but reduces the mixing to $\theta_{\tilde{\mu}} = -\pi/2+\pi/16$ ($B \simeq 469$), while the red curves correspond to a model with $\theta_{\tilde{\mu}} = -\pi/2+\pi/8$ but the trilinear coupling is reduced to $y \simeq 21$ ($B \simeq 530$). The top left panel shows the potentials near the beginning of the respective field trajectories\footnote{For both numerical stability of the FindBounce solutions and visual clarity, we have rescaled the kinetic term, potential and the Euclidean radius in Eq.~(\ref{eq:bounceA}) to be dimensionless. Note that the the bounce action we consider here is invariant under such transformations.} and each curve is normalized to the value of the potential at the TV for the corresponding model. For all models shown, we see that the bounce solutions begin trajectories fairly displaced from the TV, but that models with smaller $y$ or $\theta_{\tilde{\mu}}$ further from maximal mixing tend to have trajectories which start marginally closer to the TV. This follows from the TV being deeper for models with larger trilinear couplings or mixing angles closer to maximal and, thus, the associated bounce solutions can begin relatively further from the TV in order for the trajectories to end in the FV.  In addition, field trajectories for models with deeper minima tend to move more quickly, in terms of the Euclidean radius $\rho$, through the potential from the respective starting points of each bounce solution.

We can see the cumulative effect of the slower moving bounce solutions in the lower left panel of Fig.~\ref{fig:bsol}, which is similar to the top panel but for $\rho$ which is near the SP for each trajectory. In particular, for the bounce solution corresponding to the model with $\theta_{\tilde{\mu}} = -\pi/2+\pi/8$ and $y \simeq 21$ (red curve) the slower start at small $\rho$ can lead to a significantly larger $\rho$ at which the potential is maximized along the field trajectory. While the delay of $\Delta \rho \simeq 2$ might not seem particularly relevant for bounce solutions which extend to $\rho \simeq 270$, note that the integrand in Eq.~(\ref{eq:bounceA}) is $\propto \rho^3$ so that even such a small delay in the field trajectory can have a significant impact on the bounce action. The manifestation of these effects in the calculation of the bounce action is clearer in the right panel of Fig.~\ref{fig:bsol}, in which we have plotted the separate contributions to the integrand in Eq.~(\ref{eq:bounceA}) from $ \mathcal{T} $ and $ \mathcal{V} $, as well as the sum. After accounting for the factor of $\rho^3$ in the integrand, we see how the smuon mixing angles closer to maximal and larger $y$ ultimately yield smaller bounce actions. As mentioned above for models with deeper true minima, the contribution to the bounce action from $ \mathcal{V} $ is smaller since the field trajectory in such cases is able to start further away from the TV. If we then look to the associated contribution to the bounce action from  $ \mathcal{T} $, we see that smaller values of the potential at the beginning of the trajectory lead to smaller subsequent contributions from the kinetic energy. As a consequence, the cancellation between the two contributions when summed is more precise for models with deeper true minima and the resulting bounce action is smaller. 

\begin{figure}[t!]
\centering
\includegraphics[scale=0.54]{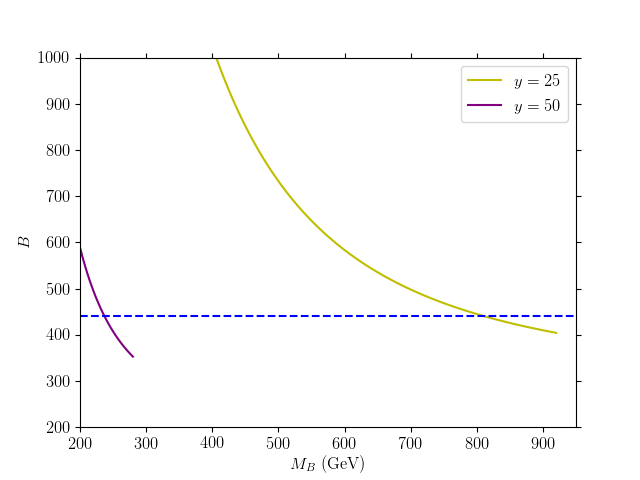} \includegraphics[scale=0.54]{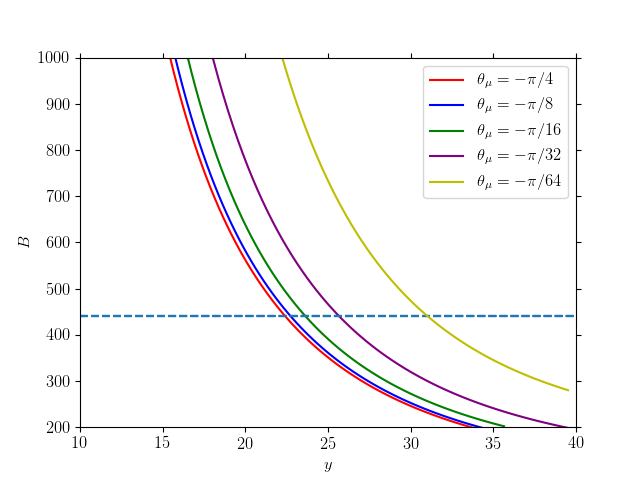}
\caption{{\sl Left panel:} Plot of the bounce action as a function of $M_B$ for the models with fixed $y$ satisfying $g_\mu -2$ and the relic density in Fig.~\ref{fig:oh2iso1}. {\sl Right panel:} Plot of the bounce action as a function of $y$ for the models with fixed smuon mixing angle satisfying $g_\mu -2$ and the relic density in Fig.~\ref{fig:oh2iso2}. The horizontal blue dashed lines show the lower limit $B > 440$, below which the EW vacuum is short-lived compared to the age of the Universe.  In both panels we fit curves to the bounce actions calculated by FindBounce at points with $200 \lesssim B \lesssim 1000$, as the calculation can become numerically unstable for models with $B \ll 440$ or $B \gg 440 $. The dependencies of the bounce actions are good fits for model points along the both of the corresponding RHB and LHB in Fig.~\ref{fig:oh2iso1} and Fig.~\ref{fig:oh2iso2}. In the right panel, we label each curve for the mixing angle, $\theta_{\tilde{\mu}}$, corresponding to points from Fig.~\ref{fig:oh2iso2} along the LHB, but the fit is also valid for points along the RHB in Fig.~\ref{fig:oh2iso2} for smuon mixing angles $-\pi/2-\theta_{\tilde{\mu}}$.
}
\label{fig:bvsy} 
\end{figure}

More generally, the bounce action increases for potentials with smaller trilinear terms until EW vacuum becomes
the global minimum of the scalar potential. For all points we consider which satisfy both the relic density and the
$g_\mu-2$, the EW vacuum is metastable. In Fig.~\ref{fig:bvsy}, we show the dependence of the bounce action on $M_{\tilde{B}} $ for points from Fig.~\ref{fig:oh2iso1} and on $y$ for points from Fig.~\ref{fig:oh2iso2}. As in Fig.~\ref{fig:bsol}, we see the bounce action decreases for models with larger $y$ and mixing angles closer to maximal. To constrain the parameter space of our simplified model, we interpolate to find the point on each curve which corresponds to a bounce action of $B = 440$. We consider all points along the curves with $B < 440$ to be excluded since the tunneling time the EW vacuum to the TV is not sufficiently large compared to the age of the Universe. For the points which satisfy $g_\mu-2$ and the relic density with larger $y$ in Fig.~\ref{fig:oh2iso1}, the constraints on $M_{\tilde{B}}$ vary significantly depending on the value of $y$, with $M_{\tilde{B}} \lesssim 800 \,$GeV for $y=25$ and $M_{\tilde{B}} \lesssim 250 \,$GeV for $y=25$. For model points from Fig.~\ref{fig:oh2iso2}, we see the constraints on the trilinear coupling vary from $y \lesssim 30$ for the mixing angles further from maximal to $y \lesssim 25$ for mixing angles closer to maximal.  When compared to the constraints from perturbative unitarity, it is clear at least for the  simplified model we have implemented that requiring a sufficiently stable EW vacuum provides for a more stringent limitation on the viable parameter space.

\section{Direct detection}  \label{sec:dd}
\noindent
In this Section we consider the prospects of testing our model by detecting the nuclear recoils induced by DM-nucleus elastic scattering. Since the Bino DM candidate in our model has no tree-level couplings to quarks, such direct detection signals are in general expected to be small, except for specific regions in parameter space in which the leading 1-loop contributions to the scattering cross section can be enhanced. We calculate the sensitivity of direct detection searches to our model within the effective field theory (EFT) framework for WIMP-nucleon scattering. 

The EFT operators relevant for pure Binos coupled to mixed-chirality sfermions (either squarks or sleptons) have been collected systematically in the literature, \eg, in Ref.~\cite{solonDD}. For our model, at the level of Bino-quark interactions, three sets of penguin diagrams contribute at leading order in perturbation theory. The loops in all of these diagrams involve the SM muon and the smuons introduced in our theory to satisfy $g_\mu -2$. Three possible states can mediate the corresponding $t$-channel interaction with the quark current: the SM Higgs, the $Z^0$ boson and the photon. These contribute, respectively, to the scalar (spin-independent), pseudo-vector (spin-dependent) and anapole operators:
\be
\mathcal{L}_{\tilde{B}^0\,q}  =   
c_q^{(0)} \bar{\tilde{B}}^0  \tilde{B}^0~m_q \bar{q}q
+ c_q^{(1)}  \bar{\tilde{B}}^0 \gamma_\mu \gamma^5 \tilde{B}^0~\bar{q}\gamma^\mu \gamma^5 q
+ e Q_q  c_A(k^2) \bar{\tilde{B}}^0 \gamma_\mu \gamma^5 \tilde{B}^0~\bar{q}\gamma^\mu q, 
\label{Binoq}
\ee
where the Wilson coefficient $c_q^{(0)}$, $c_q^{(1)}$, and $c_A(k^2)$ are obtained by computing the loops and integrating out sleptons and massive mediators (full expressions can be found, \eg, in the Appendix of Ref.~\cite{solonDD}). The scalar and anapole operators yield the dominant contribution to the scattering cross section since the separate contributions add up coherently when folded on the nucleon and then nucleus currents, with scattering amplitudes scaling respectively as the mass and atomic number of the nucleus.  Also, the Wilson coefficients for the scalar and anapole operators can be enhanced in some regions of the parameter space most relevant for our model. 

Starting with the photon-mediated anapole operator, the only operator typically considered within the MSSM for spectra with pure Bino DM and light sleptons, the expression for $c_A(k^2)$ simplifies in the limit in which the dependence on the momentum transfer can be neglected ($k^2 \ll m_\mu^2$):
\be
\label{anapolezero}
c_A \approx \frac{e}{48\pi^2}\sum_{i=1,2}\alpha_\mu^{(i)}\beta_\mu^{(i)}\int_0^1 dx \frac{3x-2}{x+(1-x)t_i-x(1-x) r_i},
\ee
where the couplings $\alpha_\mu^{(i)}$ and $\beta_\mu^{(i)}$ are obtained from rewriting the Bino-muon-smuon interaction for mass eigenstates in the form:
\be
\label{binosleplep}
\mathcal{L} \supset \sum_{i=1,2}\left\{\tilde{\mu}_i \bar{\mu}\left[\alpha_\mu^{(i)} + \beta_\mu^{(i)}\gamma_5\right] \tilde{B}^0 + \text{h.c.}\right\},
\ee
and $r_i \equiv {M_{\tilde{B}}^2}/{M_{\tilde \mu_i}^2}$ and $t_i \equiv {m_\mu^2}/{M_{\tilde \mu_i}^2}$. To estimate the integral in Eq.~(\ref{anapolezero}), one can perform an expansion at $t_i \ll 1$  and $r_i \rightarrow 1$ or $r_i \rightarrow 0$ to find:
\be
L_A(r_i,t_i) \equiv \int_0^1 dx \frac{3x-2}{x+(1-x)t_i-x(1-x) r_i} \approx \begin{cases}
\left(2 - \frac{2}{t_i}-3 \ln t_i\right) + \left(4 +\frac{1}{t_i^2}-\frac{5}{t_i}-3\ln t_i\right)(1-r_i)+ \mathcal{O}((1-r_i)^2),\\
~\\
\left(3 - 3t_i + 2\ln t_i\right) + \left(\frac{7}{2} - 5t_i + \frac{3t_i^2}{2} + 2\ln t_i\right)r_i + \mathcal{O}(r_i^2).
\end{cases}
\ee
We can see that as one smuon becomes nearly degenerate in mass with the Bino---the relevant regime for coannihilations in the early Universe---the anapole moment is enhanced due to the large hierarchy between the smuons and the muon; in the regime of large smuon-Bino mass splittings there is instead only a mild logarithmic enhancement. 

Regarding the Wilson coefficient for scalar interactions $c_q^{(0)}$, while the small muon Yukawa coupling suppresses the contribution from Higgs mediation with the muon in the loop, an enhancement can be present if the Higgs-smuon coupling $y_{H_2^0 ij}$ (see Eq.~(\ref{yh0ij})) is large. In the limit of a massless muon, one has~\cite{solonDD}:
\be
c_q^{(0)} \simeq \frac{g^2}{16\pi^2 M_{H^0}^2 M_{\tilde{B}}} \sum_{i\leq j} 
(\alpha_\mu^{(i)}\alpha_\mu^{(j)}+\beta_\mu^{(i)}\beta_\mu^{(j)}) y_{H_2^0 ij} \, \frac{r_i}{1 - r_i/r_j}
\int_0^1 dx (1-x) \ln\left(\frac{1/r_i-x}{1/r_j-x}\right)\,.
\ee
The computation of the integral can be performed analytically and the largest contribution in the limit $r_2\ll r_1$ arises from the case $i=j=1$,
\be
c_q^{(0)} \simeq \frac{g^2\,y_{H_2^0 11}}{32\pi^2 M_{H^0}^2 M_{\tilde{B}}} 
\left(\lambda_{\tilde \mu_R}^2 \sin^2(\theta_{\tilde{\mu}}) + \lambda_{\tilde \mu_L}^2 \cos^2(\theta_{\tilde{\mu}})\right)
\left[1+\frac{1-r_1}{r_1} \ln(1-r_1)\right]\,,
\ee
with the Higgs coupling that, at large $y$ and sizable left-right smuon mixing, tends to $y_{H_2^0 11}\rightarrow y\,|\sin(2 \theta_{\tilde{\mu}})|$, and hence can potentially lead to a $y^2$ scaling of the scattering cross section.

The computation of the scattering rate on a nucleus proceeds with the standard steps. First the Lagrangian in Eq.~(\ref{Binoq}) is folded on nucleon states (protons and neutrons), taking into account QCD nucleon form factors, to find the EFT for DM-nucleon interactions. Then a nonrelativistic reduction is performed, finding:
\begin{eqnarray}
\mathcal{L}_{NREFT} = \sum_{N=p,n} \left[c_N^{(0)} \mathcal{O}_1^{(N)} - 4 c_N^{(1)} \mathcal{O}_9^{(N)}\right] - e c_A(k^2) \left[2 \mathcal{O}_8^{(p)} - 2 \mathcal{O}_9^{(p)}\right],
\end{eqnarray}
where $\mathcal{O}_1$ is just the identity operator acting on isospin space, while
\begin{eqnarray}
\mathcal{O}_8 \equiv \vec{S}_\chi \cdot \vec{v}^\perp,\quad \mathcal{O}_9 \equiv i \vec{S}_\chi \cdot \left(\vec{S}_N \times \frac{\vec{k}}{m_N}\right),
\end{eqnarray}
where $\vec{S}_N$ and $\vec{S}_\chi$ are, respectively, the spin of the nucleon $N$ and the spin of the DM species $\chi$, and $\vec{v}^\perp$ is the component of the DM-nucleon relative velocity orthogonal to the momentum transfer $\vec{k}$. The coefficients $c_N$ are obtained as a sum over quark flavors for the Wilson coefficients $c_q$, weighted by the associated nucleon form factors. Having performed the reduction on this operator basis, we can use a generic tool such as \verb|DDCalc| \cite{Workgroup:2017lvb,Athron:2018hpc} to calculate the recoil spectrum for a given nucleus.

Now that we have the necessary ingredients to compute the nuclear recoil spectrum, we can address the potential sensitivity of direct detection searches to our model. To check whether a model is excluded, one must compute the test statistic $\lambda_{TS}$, defined as
\begin{eqnarray}
\label{tsdef}\lambda_{TS} \equiv -2 \ln \frac{\mathcal{L}(N_o = 0,b\vert N_p)}{\mathcal{L}(N_o, b\vert N_p)},
\end{eqnarray}
where the likelihood function is a Poisson distribution given by
\begin{eqnarray}
\mathcal{L}(N_o,b\vert N_p) = \frac{(b+N_p)^{N_o}}{N_o!}e^{-(b+N_p)},
\end{eqnarray}
$b$ is the number of background events, $N_p$ is the number of expected events, and $N_o$ is the observed number of recoil events. Then the criterion for obtaining the region of the parameter space that is rejected at 90\% CL is 
\begin{eqnarray}
\label{ts90cl}\lambda_{TS} \leq -1.64,
\end{eqnarray}
which follows from the fact that $\lambda$ follows a half-chi squared distribution. $N_p$ depends on the model parameters. We use the DDCalc package \cite{GAMBITDarkMatterWorkgroup:2017fax,GAMBIT:2018eea} to compute both the event rates (including all interaction terms at 1-loop level) and the likelihood functions for a particular detector.

\begin{figure}[ht]
\centering
\begin{tabular}{cc}\hline
RHB&LHB\\\hline\hline
\includegraphics[scale=0.5]{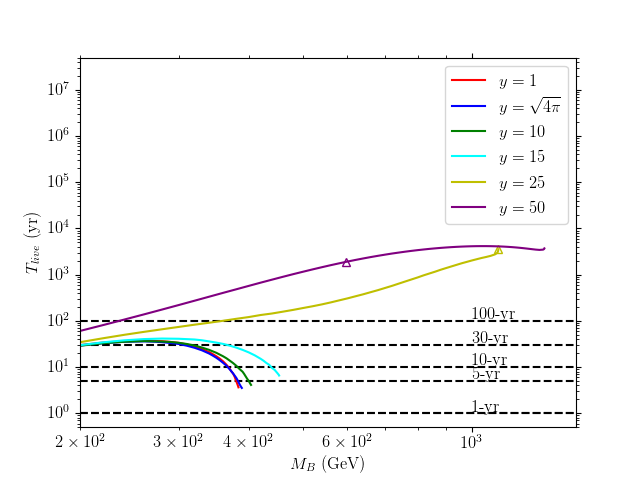}&\includegraphics[scale=0.5]{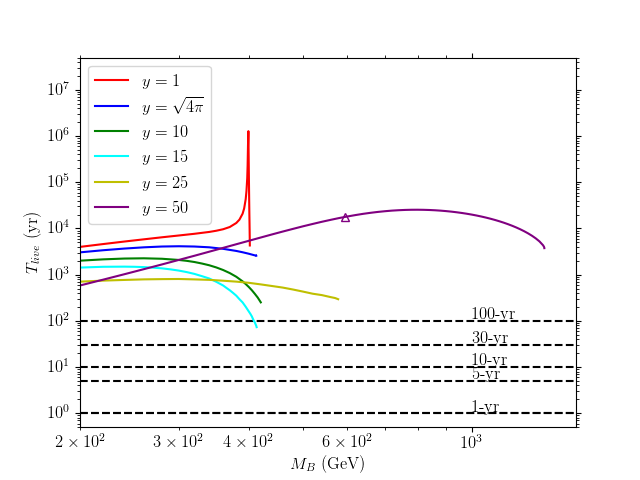}\\
(a)&(b)\\
\includegraphics[scale=0.5]{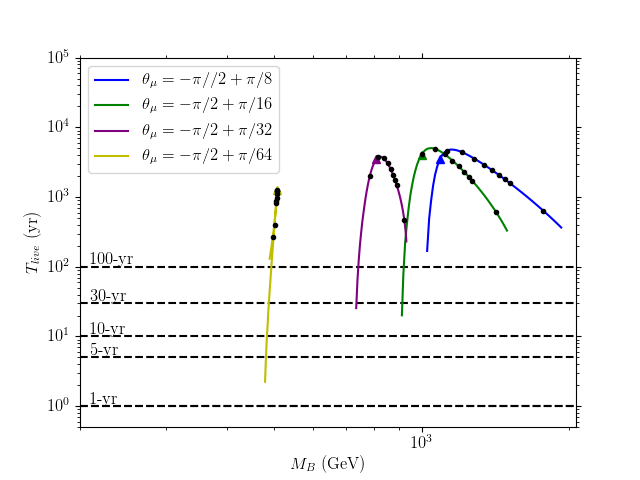}&\includegraphics[scale=0.5]{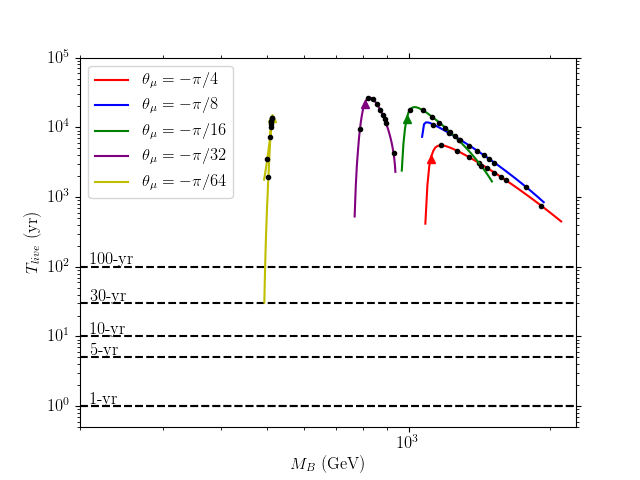}\\
(c)&(d)
\end{tabular}
\caption{Plots of the $T_{live}$, the exposure time necessary for a DARWIN-like detector with a fiducial target mass of $\unit[40]{t}$ to be sensitive at 90\% CL to scattering cross sections for models satisfying $g_\mu -2$ and the relic density. The panels in the top row correspond to model points along the RHB (a) and LHB (b) shown in Fig.~\ref{fig:oh2iso1} with fixed $y$, while the panels in the bottom row correspond to model points along the RHB (c) and LHB (d) shown in Fig.~\ref{fig:oh2iso2} with fixed $\theta_{\tilde{\mu}}$. For panels (c) and (d), the black dots correspond to those in  Fig.~\ref{fig:oh2iso2} with $y = 30, 40, 50, 60, 70, 80, 90, 100, $ and $200$ as $M_{\tilde{B}}$ increases along each respective curve. The triangular markers along various curves in all of the panels indicate the transition from an anapole-dominated recoil spectrum to a Higgs-dominated one. The scattering cross sections for models along those curves without triangles are dominated by anapole interactions.}
\label{fig:fixthexpose} 
\end{figure}

We focus on DARWIN~\cite{Schumann:2015cpa} as a benchmark for a future detector, and check what exposure time would be necessary to be sensitive at 90\% CL to the nuclear recoil spectra associated with our model. For a given set of model parameters, this check gives an estimate of how close to (or far from) an eventual discovery that a hypothetical DARWIN detector would be. Performing the reverse, \ie~projecting the sensitivity of the detector onto the parameter space of the model, is more subtle, since different operators can contribute to the scattering cross section: Although in general anapole interactions are the most relevant, the other operators cannot be neglected, and scalar interactions can actually become dominant at large $y$ and sizable mixing. We compute the quantity $T_{live}$, defined as the minimal required live time\footnote{While $T_{live}$ is extrapolated to extremely large values to demostrate the challenge of probing our model with direct detection, note that this estimate obviously does not account for the practical implications of such large exposure times or other relevant effects. For example, at large enough exposure the sensitivity of any direct detection experiment would become limited by the atmospheric and solar neutrino background \cite{Billard:2013qya,OHare:2016pjy}. } for a DARWIN-like detector with a fiducial target mass of $\unit[40]{t}$~\cite{aalbers2016darwin}, to reach $\lambda_{TS} = -1.64$. We focus on the parts of parameter space most relevant for models satisfying $g_\mu -2$ and the relic density shown in Figs.~\ref{fig:oh2iso1} and \ref{fig:oh2iso2}. 

In Fig.~\ref{fig:fixthexpose}, $T_{live}$ is plotted versus the Bino mass $M_{\tilde{B}}$ for each of the curves in each of the panels  from Figs.~\ref{fig:oh2iso1} and \ref{fig:oh2iso2}. The left panels show $T_{live}$ for points along the RHB and the right panels for points along the LHB; the top panels correspond to points from the curves in Fig.~\ref{fig:oh2iso1} with fixed $y$ and the bottom panels correspond to points from the curves in Fig.~\ref{fig:oh2iso2} with fixed $\theta_{\tilde{\mu}}$. We see that $T_{live}$ for most of the points that satisfy $g_\mu-2$ and the relic density is above the 5-year run time foreseen for DARWIN, except for a subset of points in Panel (a) with $y \leq 10$ and a few points in Panel (c) corresponding to $\theta_{\tilde{\mu}} = -\pi/2+\pi/64$. The improvement in sensitivity for points along the RHB relative to the LHB can be explained in part by the gauge couplings associated with our choices of Bino-muon-smuon couplings, $\vert Y_R\vert = 2\vert Y_L\vert$. Also, in models with mixing angles closer to $\theta_{\tilde{\mu}} = 0$ and $y$ sufficiently small, the sneutrino can be degenerate enough in mass with the Bino such that coannihilation processes involving the sneutrino become relevant for depleting the relic density. For these points along the LHB, the (mostly left-handed) lightest smuon can be heavier than the (mostly right-handed) lightest smuon for the corresponding points along the RHB. Thus, the contribution from the anapole moment to the scattering cross section, which is dominant in these cases, can be relatively suppressed along the LHB.

In Panels~(a) and~(b) we also see that the exposure time monotonically decreases for Bino masses larger than $\unit[300]{GeV}$ for points with $y \leq 10$. As shown in Fig.~\ref{fig:oh2iso1}, these models correspond to points with relative mass splittings between the Bino and lightest smuon $\sim 10^{-3}-10^{-2}$, where the required mass splitting decreases with increasing $M_{\tilde{B}}$. The associated trend in $T_{live}$ is consistent with the discussion above regarding the contribution to the scattering cross section from anapole interactions, which sharply increases in the limit of small mass splitting between Bino and lightest smuon. On the other hand, models with sizable mixing angles in Panels (c) and (d) exhibit a ``turnaround" in $T_{live}$ as $M_{\tilde{B}}$ increases. Referring to Fig. \ref{fig:oh2iso2}, we can see this turnaround is the result of two effects: at fixed $\theta_{\tilde{\mu}}$ both $y$ and the mass splitting between the Bino and lightest smuon must increase to satisfy $g_\mu-2$ and the relic density for larger $M_{\tilde{B}}$. The latter suppresses anapole interactions, while a large $y$ and sizable mixing enhances Higgs mediated scalar interactions, with the contribution to the recoil spectrum from Higgs exchange becoming dominant over the anapole contribution. To show where the Higgs exchange starts to dominate the scattering cross section in Fig.~\ref{fig:fixthexpose}, we indicate with a triangle along the relevant curves where the anapole contribution becomes subdominant for increasing $M_{\tilde{B}}$.

\begin{figure}[t]
\centering
\includegraphics[scale=0.42]{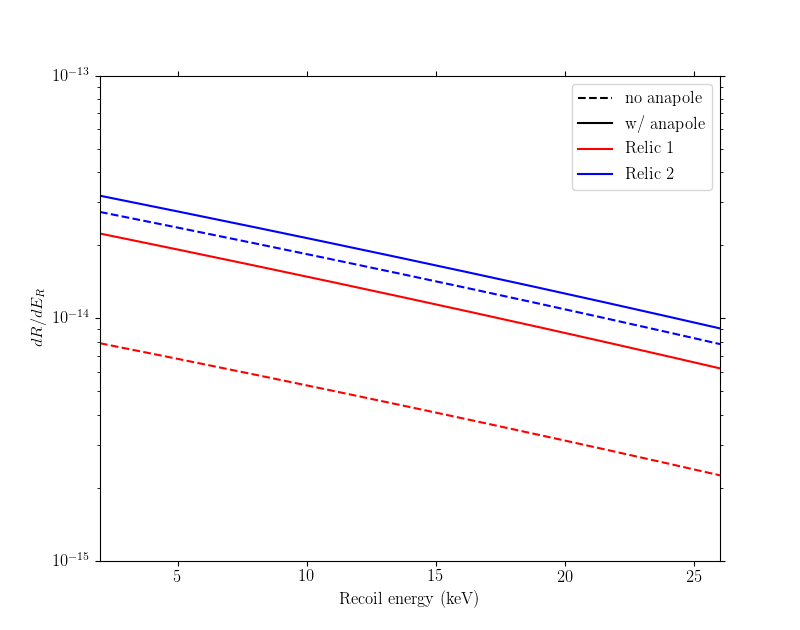} \includegraphics[scale=0.42]{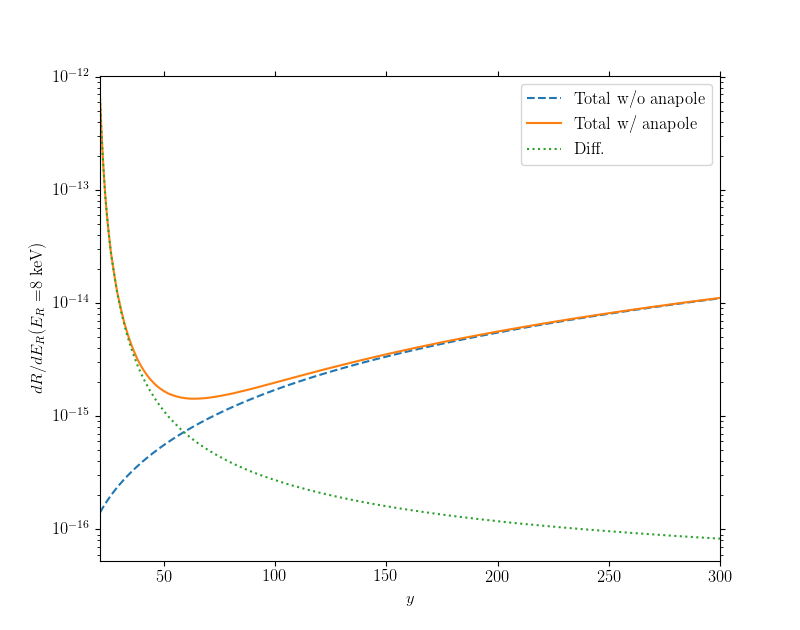}
\caption{{\sl Left panel:} Recoil spectra for two benchmark points, either including all contributions to the scattering cross section (solid lines) or excluding the contribution from anapole interactions (dashed). ``Relic 1" and ``Relic 2" correspond to benchmark points along the $\theta_{\tilde{\mu}} = -\pi/2+\pi/64$ curve from Fig.~\ref{fig:oh2iso2} with $y=50$ and $y=100$, respectively. {\sl Right panel:} Normalization of the differential recoil rate at $E_R = \unit[8]{keV}$ versus $y$ along the $\theta_{\tilde{\mu}} = -\pi/2+\pi/64$ curve from Fig.~\ref{fig:oh2iso2}, either including (solid line) or excluding (dashed) the contributions from anapole interactions. The difference between the normalizations of the recoil spectra with or without anapole contributions is shown by the dotted curve.}
\label{fig:spex} 
\end{figure}

In order to see the relative contributions to the the recoil spectra from either anapole interactions or Higgs (and $Z^0$) exchange in more detail, we show the recoil spectra for two benchmark points in the left panel of Fig.~\ref{fig:spex}. Recoil spectra labelled ``Relic 1" and ``Relic 2" correspond to benchmark points along the $\theta_{\tilde{\mu}} = -\pi/2+\pi/64$ curve from Fig.~\ref{fig:oh2iso2} with $y=50$ and $y=100$, respectively. For each case, we display the recoil spectrum assuming that either all Higgs, $Z^0$, and photon exchange processes or all processes except photon exchange contribute to the scattering cross section. We observe that the anapole contribution dominates the recoil spectrum for Relic 1, while it is subdominant for the case of Relic 2 where $y$ is larger. The right panel of Fig.~\ref{fig:spex} shows the normalization of the recoil spectrum at $E_R = \unit[8]{keV}$, along the $\theta_{\tilde{\mu}} = -\pi/2+\pi/64$ curve from Fig.~\ref{fig:oh2iso2} either including or excluding the contributions from anapole interactions. As discussed above, the relative mass splitting between the Bino and the lightest smuon generally increases for larger $y$ along the curves of constant $\theta_{\tilde{\mu}}$ in Fig.~\ref{fig:oh2iso2}. The increased mass splitting suppresses the anapole contribution to the total recoil spectrum. Also, increasing $y$ simply drives up the Higgs-slepton trilinear coupling, resulting in a larger non-anapole contribution to the recoil spectrum.

\section{Conclusions} \label{sec:con}
\noindent
Rather generic extensions to the SM of particle physics can provide extra 1-loop contributions to the muon $g-2$, possibly accounting for the 4.2$\sigma$ anomaly reported by the E989 experiment. The embedding of a dark matter candidate in such extensions has also been discussed on rather general grounds. In this work, we have considered a minimal BSM framework in which the extra states responsible for the muon $g-2$ discrepancy also provide for a dark matter candidate and determine its relic abundance in the early Universe. 

The analysis has been carried out within a specific model in which the essential BSM states are: a Majorana fermion with no electric charge or muonic lepton number, playing the role of dark matter; a scalar with mixed chirality carrying electric and muon leptonic charges. The particle spectrum of this model, with the appropriate choices of $SU(2)_L \times U(1)_Y$ quantum number assignments, maps onto a small subset of the particle content of the MSSM, from which we have borrowed the terminology---the extra states just mentioned are referred to as, respectively, the Bino and a smuon---and that we exploit as an embedding framework when specifying the origin of the chiral mixing for the leptonic scalar.

The model has a reduced parameter space, essentially only 3 masses and one mixing angle. Requiring that the model satisfies the $g_\mu-2$ anomaly foliates this parameter space along left-handed or right-handed branches for the lightest smuon. Along these branches the level of left-right mixing, dictated by the chirality flip necessary for the BSM contribution to $g_\mu-2$, is much larger than what is usually considered in the MSSM under the assumption of minimal flavor violation. The phenomenology of our model then clearly departs from what is usually discussed in the context of MSSM parameter scans.

The requirement that the Bino relic density matches the dark matter density of the Universe leads us to consider scenarios in which the sleptons are just slightly heavier than the Bino (relative mass splittings of order 10\% or lower). Since the sleptons interact with the heat bath more efficiently than the Bino, the charged scalars can drive thermal freeze out via coannihilation effects. For coannihilating particles with relatively light masses $\lesssim 400 \,$GeV, the parameter space for which our model satisfies both $g_\mu-2$ and the relic density is similar to that of Bino-slepton coannihilation scenarios previously investigated in the so-called bulk region of the MSSM. However, once the assumption of minimal flavor violation is relaxed, the viable parameter space of our model opens up into regions in which the coannihilating particles are sensibly heavier. We find that a proper description of this effect is given in terms of the dimensionless parameter $y$, introduced in Eq.~(\ref{eq:y}), which is a measure of the mass splitting between the lighter and heavier smuons relative to the weak scale, weighted by the left-right mixing angle. For moderate values of $y$, we move away from the usual slepton coannihilation regime in the bulk region of the MSSM, to scenarios with coannihilating particle masses at the TeV scale and beyond. 

Upon a detailed examination of the relic density calculation in this parameter space characterized by heavy Binos, large $y$ and sizable mixing, we see that some of the relevant cross sections tend to become large, although not large enough to violate face-value perturbative unitarity bounds. Taking one step further and borrowing the structure of the full scalar potential from the MSSM, a comprehensive analysis of the full S-matrix shows that unitarity rules out large to moderate values of $y$, depending on whether the mixing is mild or maximal. The parameter space is constrained even further when considering the stability of the electroweak vacuum. For models with sizable smuon mixing and large trilinear couplings, the scalar potential can develop minima deeper than the EW vacuum. Requiring that the tunneling time from the EW vacuum to the true vacuum is longer than the age of the Universe sets the tightest constraints on the parameter space of the model: $y$ cannot exceed moderate values regardless of the smuon mixing angle and the Bino mass scale cannot be larger than about \unit[1]{TeV}.

The prospects of testing our scenario with the next generation of direct detection experiments are unfortunately limited to a marginal portion of the viable parameter space. There is no tree-level coupling between the Bino and SM quarks in our model, and the anapole operator relevant for direct detection searches is only sufficiently enhanced for cases with very small mass splittings between the Bino and lightest smuon. On the other hand, a future lepton collider with a relatively large center of mass energy could directly probe the extended parameter space of our model. Since the most stringent constraints arise from perturbative unitarity and vacuum stability in our simplified model, it would also be interesting to consider the phenomenological implications of embedding our simplified model into a framework which provides for a more theoretically consistent extension of the SM.
\section*{Acknowledgments}
This work was supported by the research grant ``The Dark Universe: A Synergic Multimessenger Approach" number 2017X7X85K under the program PRIN 2017 funded by the The Italian Ministry of Education, University and
Research (MIUR), and by the European Union's Horizon 2020 research and innovation program under the
Marie Sklodowska-Curie grant agreement No 860881-HIDDeN. JTA gratefully acknowledges the hospitality and support of the International Centre for Theoretical Physics (ICTP). PS would like to thank Sebastian Baum for helpful discussions.
\bibliographystyle{kp.bst}
\bibliography{ASU_paper-final}
\end{document}